\documentclass[12pt,preprint]{aastex}

\newcommand\iso[2]{$^{\rm #1}$#2}

\def\kmsec{\mbox{km~s$^{\rm -1}$}}
\def\teff{\mbox{$T_{\rm {eff}}$}}
\def\logg{\mbox{log $g$}}
\def\vt{\mbox{v$_{\rm t}$}}
\def\BmV0{\mbox{(B-V)$^{\rm o}$}}
\def\VmK0{\mbox{(V-K)$^{\rm o}$}}
\def\MV0{\mbox{M$_{\rm V}^{\rm o}$}}

\def\etal{\mbox{et al.}}
\def\eg{\mbox{e.g.}}
\def\ie{\mbox{i.e.}}

\def\third{{3$^{\rm rd}$}}
\def\deg{{$^{\circ}$}}
\def\bd17{\mbox{BD +17\deg 3248}}
\def\cs22{\mbox{CS 22892-052}}
\def\gtaprx{ \mathrel{  \vcenter{
                        \offinterlineskip \hbox{$>$}
                        \kern 0.3ex \hbox{$\sim$}    } } }
\def\ltaprx{ \mathrel{  \vcenter{
                        \offinterlineskip \hbox{$<$}
                        \kern 0.3ex \hbox{$\sim$}    } } }
\begin{document}

\title{HST Observations of Heavy Elements in 
Metal-Poor Galactic Halo Stars}

\author{
John J. Cowan,\altaffilmark{1}
Christopher Sneden,\altaffilmark{2}
Timothy C. Beers,\altaffilmark{3}
James E. Lawler,\altaffilmark{4}\\
Jennifer Simmerer,\altaffilmark{5}
James W. Truran,\altaffilmark{6}
Francesca Primas,\altaffilmark{7}
Jason Collier,\altaffilmark{8}
and
Scott Burles\altaffilmark{9} \\
}

\altaffiltext{1}{Department of Physics and Astronomy,
University of Oklahoma, Norman, OK 73019; cowan@nhn.ou.edu}
\altaffiltext{2}{Department of Astronomy and 
McDonald Observatory, University of Texas, 
Austin, Texas 78712; chris@verdi.as.utexas.edu}
\altaffiltext{3}{
Department of Physics and Astronomy and JINA: Joint Institute for
Nuclear Astrophysics, Michigan State
University, East Lansing, MI 48824; beers@pa.msu.edu}
\altaffiltext{4}{Department of Physics, University of Wisconsin, Madison,  
Madison, WI 53706; jelawler@wisc.edu}
\altaffiltext{5}{Department of Astronomy and McDonald Observatory, 
University of Texas, 
Austin, Texas 78712; jensim@astro.as.utexas.edu}
\altaffiltext{6}{
Department of Astronomy and Astrophysics, Enrico Fermi
Institute, University of Chicago, Chicago, IL 60637;
truran@nova.uchicago.edu}
\altaffiltext{7}{
European Southern Observatory, Karl-Schwarzschild Strasse 2,
D-85748 Garching bei Muenchen; fprimas@eso.org}
\altaffiltext{8}{Department of Physics and Astronomy,
University of Oklahoma, Norman, OK 73019; collier@nhn.ou.edu}
\altaffiltext{9}{
Department of Physics, Massachusetts Institute of Technology,
Cambridge, MA 02139; burles@mit.edu}

\begin{abstract}
We present new abundance determinations 
 of neutron-capture elements Ge, Zr, Os, Ir, 
and Pt in a sample of 11 metal-poor (--3.1 $\le$ [Fe/H] $\le$ --1.6) 
Galactic halo giant stars, based on Hubble 
Space Telescope UV and Keck~I optical high-resolution spectroscopy. 
The stellar sample is dominated by $r$-process-rich stars such as the
well-studied CS~22892-052 and \bd17, but also includes  the 
$r$-process-poor, bright giant HD~122563.
Our results demonstrate that abundances of 
the \third\ $r$-process peak elements 
Os, Ir and Pt in these metal-poor halo stars 
are very well-correlated among themselves, and with 
the abundances of the canonical $r$-process element Eu (determined in other
studies), thus arguing for a common origin or 
site for $r$-process nucleosynthesis of heavier (Z~$>$~56) elements.
However,  the large (and correlated) scatters of [Eu,Os,Ir,Pt/Fe] 
suggests that the heaviest neutron-capture $r$-process elements are not 
formed in {\it all} supernovae.
In contrast, the Ge abundances of all program stars track their 
Fe abundances, 
very well. 
An explosive process on iron-peak nuclei ({\it e.g.}, the $\alpha$-rich 
freeze-out in supernovae), rather than neutron 
capture, appears to  have been the dominant synthesis mechanism for 
this element at low metallicities -- Ge abundances seem completely
uncorrelated with Eu.
The  
correlation (with very small scatter) of Ge and Fe abundances
suggests that Ge must have been produced rather commonly in stars --
even at early times in the Galaxy -- over a wide range of metallicity.
The Zr abundances show much the same behavior as Ge with (perhaps)
somewhat more scatter, suggesting some variations in abundance with respect
to Fe.
The Zr abundances also do not vary cleanly with Eu abundances, indicating a 
synthesis origin different than that of heavier neutron-capture elements.

Detailed abundance distributions, for \cs22\ and \bd17, combining the 
new elemental determinations for Os-Pt and recently published Nd and Ho 
measurements, show excellent agreement with the solar system $r$-process 
curve from the elements Ba to Pb. 
The lighter $n$-capture elements, including Ge, in general fall below the 
same solar system $r$-process curve that matches the heavier elements.  

\end{abstract}


\section{Introduction}

Elemental abundances in metal-poor Galactic halo stars are providing 
evidence of the earliest Galactic nucleosynthesis history and clues 
about the identities of the first stellar generations, the progenitors 
(or predecessors) of 
the halo stars. 
The neutron-capture ($n$-capture) elements -- formed in slow ($s$-process)
and rapid ($r$-process) neutron-capture nucleosynthesis -- were synthesized 
in these first stars and later ejected into the interstellar medium and
eventually incorporated into the halo stars (see recent reviews by 
Cowan \& Thielemann 2004; Truran \etal\ 2002; Sneden \& Cowan 2003; 
Cowan \& Sneden 2004).

The focus of $n$-capture abundance studies in metal-poor stars has been to 
examine correlations among these elements in an attempt to help
to identify the sites of nucleosynthesis for these elements and also to 
trace their evolution with Galactic metallicity.
Most of the previous studies in this field have focused on the rare-earth 
elements ({\it e.g.}, Ba and Eu) that are easily detectable with 
ground-based telescopes.  
Such studies have indicated a consistency, or scaling, (for 
$r$-process-rich stars) between the total $n$-capture elemental 
abundances and the solar system $r$-process-only abundances.  
Until recently, however,  it has not been possible to examine the full range of
the $n$-capture elements, as many of them have dominant transitions in
the near UV that are accessible only to telescopes such as the Hubble 
Space Telescope (HST). 
With the Space Telescope Imaging Spectrograph (STIS), it is now possible to 
make reliable abundance determinations of the heaviest stable $n$-capture 
elements (such as Pt) to determine whether the relative scaling of the 
rare-earth elements extends through the \third $r$-process peak, and also 
to examine some of the lighter, less-investigated elements such as Ge. 

In this paper we report the first large-sample abundance determinations of
the abundances for the elements Ge, Zr, Os and Pt using HST, supplemented 
by Keck~I High Resolution Echelle Spectrograph (HIRES) observations of the 
very heavy $n$-capture element Ir, in a group of 11 metal-poor halo stars.  
In \S2 we present the spectroscopic data, in \S3 we describe the
abundance analyses, and in \S4 we discuss the correlations among the 
elements and the implications for stellar nucleosynthesis and 
Galactic chemical evolution. 
In this section we also revisit the detailed elemental abundance 
distributions for the two well-studied $r$-process-rich stars \cs22 and 
\bd17, followed by our conclusions in \S5.

\section{Observations and Reductions}
                                     
High-resolution ultraviolet spectra were gathered with the Hubble Space
Telescope Imaging Spectrograph (HST/STIS).
The instrumental setup included the echelle grating E230M centered at
2707~\AA, an entrance aperture of 0.2$\arcsec\times$0.06$\arcsec$,
and the near-UV MAMA detector.
These components yielded spectra in the wavelength range
2410~\AA~$\leq$~$\lambda$~$\leq$~3070~\AA\ with a spectral resolving
power of R~$\equiv$~$\lambda/\delta\lambda$~$\simeq$~30,000.
One to five individual spectra were obtained during the HST visit for
each program star, depending on the target brightness.
The observing sequence was a standard routine that included target
acquisition, peakup, and integration(s).
No special calibration exposures were needed.

Standard HST pipelines produced the one-dimensional, flat-fielded,
wavelength-calibrated spectra from the individual stellar integrations.
Algorithms implemented in the IRAF\footnote{
IRAF is distributed by the National Optical Astronomy Observatory, which
is operated by the Association of Universities for Research in Astronomy,
Inc., under cooperative agreement with the National Science Foundation.}
software system were then used to reformat these spectrum files
into standard FITS spectral images, and to combine the integrations
to produce final spectra.
The signal-to-noise ratios (S/N) of the spectra were difficult
to estimate accurately because of the strong-lined nature of the stars
in the UV wavelength region.
From comparisons of synthetic and observed spectra we estimate that
S/N~$\gtrsim$~50 in the spectral regions of interest.

We supplemented these data with visible-region high resolution spectra
obtained with the Keck~I HIRES (Vogt \etal\ 1994).
This instrument was configured to produce complete spectral coverage
in interval 3150~\AA~$\leq$~$\lambda$~$\leq$~4600~\AA\ at a resolving
power of R~$\simeq$~45,000, and 30~$\lesssim$~S/N~$\lesssim$~200 (a steadily
increasing function with increasing wavelength for these cool target
stars,  which were  with a CCD system not optimized for near-UV response).
We obtained auxiliary lamp spectra for flat-fielding and wavelength
calibration of the spectra.
The echelle extractions, including flat-field and bias correction,
cosmic-ray rejection, sky subtraction, and wavelength and flux calibration,
were accomplished with the software package MAKEE (e.g., Barlow \& Sargent
1997).

\section{Abundance Analysis}

\subsection{Model Stellar Atmospheres}

For the stars CS~22892-052 and BD~+17$^{o}$3248, recent and detailed 
analyses have been performed by Sneden \etal\ (2003) and Cowan \etal\ (2002);
their parameters have been adopted here.  
The stars HD~6755, HD~115444, HD~122563, HD~122956, HD~186478, and HD~221170  were 
analyzed with the stellar models used by Simmerer \etal\ (2004). 
The remaining stars (HD~6268, HD~1265887, and HD~175305) were 
assigned stellar models based on methods used in Simmerer \etal\
The details of these parameter selection methods are outlined in that 
paper and are summarized here.

The grid of Kurucz stellar atmosphere models with no convective overshoot 
(Castelli \etal\ 1997) were employed.
We interpolated these to the set of stellar parameters using software 
provided by A. McWilliam (1990, private communication) and I. Ivans (2002, 
private communication).  

The stellar parameters (effective temperature \teff, surface gravity \logg, 
metallicity as represented by [Fe/H], and microturbulence \vt) were then 
checked against standard spectroscopic constraints.  

\begin{itemize}
\item \teff: When possible, effective temperatures were taken from Alonso 
\etal\ (1999,2001), who give values for their calibration stars (which include 
HD~6755, HD~122563, HD~122956, HD~186478, and HD~221170).
These temperatures were then checked against the spectroscopic constraint 
that the equivalent widths (EWs) of \ion{Fe}{1} lines be uncorrelated with 
the excitation potentials of the lines.  
The Alonso \etal\ temperatures for those stars were all consistent 
with that requirement.  
Effective temperatures were calculated for the remaining stars (HD~6268, 
HD~115444, HD~126587, and HD~175305) with the Alonso \etal\ calibrations 
for infrared color indices. 
The IR colors were taken from the 2MASS database, as transformed to a system
consistent with Alonso \etal\ (see Simmerer \etal\ section 3.3.1).

\item Surface gravity: \logg\ may be derived from the standard relation 
involving \teff, absolute magnitude, and mass (\eg, Simmerer \etal\ 2004).
Most giant stars are too distant to have a well determined parallax (and 
hence absolute magnitude), so we used the M$_V$ derived by Anthony-Twarog \& 
Twarog (1994) for HD~122563, HD~122956, HD~186478, and HD~221170.  
The surface gravity for HD~6775 was calculated from its HIPPARCOS parallax 
(Perryman \etal\ 1997), as this star is close enough to have a 
well-determined value (though the HIPPARCOS distance is consistent 
with that of Anthony-Twarog \& Twarog).  
The derived gravities were then checked spectroscopically by requiring 
that the abundances of Fe from \ion{Fe}{1} and \ion{Fe}{2} be essentially 
equal (see Thevenin \& Idiart 1999). 
The Anthony-Twarog \& Twarog values for these stars all met that criterion.  
The remaining four stars also had distance estimates from 
Anthony-Twarog \& Twarog (1994), which we used to calculate surface gravities.  

\item Metallicity: A small list of \ion{Fe}{1} and \ion{Fe}{2} lines was used 
to check the derived stellar effective temperatures and surface gravities.  
This list was also used to assess the [Fe/H] metallicity.  
As per the discussion in Thevenin \& Idiart (1999), the Fe ionization 
equilibrium may not be correctly described by the LTE Saha formula.
We therefore report both [\ion{Fe}{1}/H] and [\ion{Fe}{2}/H] abundances 
in Table~\ref{tab1}.  
Note that Thevenin \& Idiart considered relatively high-gravity stars,
and a thorough investigation of nLTE effects in low gravity, low metallicity
stars has yet to be published.
Examples of attempts in this area include those of Gratton \etal\ (1999)
and Korn (2004).
These authors suggest that nLTE departures in abundances for stars of 
interest here may not be large, but the uncertainties in such analyses 
do not permit definitive conclusions to be reached on this issue.

The metallicity of the model atmosphere usually was set 0.25~dex higher 
than the derived Fe abundance to account for additional opacity from the 
$\alpha$-capture elements (whose abundances are enhanced at low metallicities). 
The [X/Fe] ratios in Table~\ref{tab3} are calculated with [\ion{Fe}{2}/H].  
For HD~6268, HD~115444, HD~1265887, and HD~175305, we used the [Fe/H] 
reported by Burris \etal\ (2000) and then applied the offsets found by 
Simmerer \etal\ (2004) to recover \ion{Fe}{1} and \ion{Fe}{2} abundances.  
This results in an offset of +0.04 dex (for \ion{Fe}{2}) and -0.08 dex 
(for \ion{Fe}{1}) from the [Fe/H] found in Burris \etal\ (2000).  

\item Microturbulence: \vt\ was set be requiring that log $(EW/\lambda)$ be 
uncorrelated with the abundance for \ion{Fe}{1} lines.  
Microturbulence values for the four stars not included in Simmerer \etal\ 
(2004) were set to 2.0~\kmsec, a typical value for stars in this 
\teff, \logg\ domain.

\end{itemize}

\subsection{Transition Data}

In cool stellar atmospheres germanium, osmium, iridium and platinum
can only be detected in their neutral species, whose strongest
transitions occur in the UV spectral region ($\lambda$~$<$~3500~\AA).
Such UV lines are the sole abundance indicators of these elements in
metal-poor stars. 
Spectra in this wavelength domain are extremely complex mixes of overlapping 
atomic and molecular features.
No completely unblended features of \ion{Ge}{1}, \ion{Os}{1},
and \ion{Pt}{1} exist at the spectral resolution of our HST/STIS data.
Therefore synthetic spectrum computations were employed in all of the
abundance determinations of this paper.

Construction of atomic and molecular line lists has been discussed in
several of our previous papers; see Sneden \etal\ (1996, 2003)
and Cowan \etal\ (2002) for overviews and application to $r$-process-rich
metal-poor giant stars.
Sneden \etal\ (1998) describe in detail and show illustrative spectra of
several of the UV features that are employed here.
Briefly, we combined laboratory data for the lines of interest with
data for other atomic and molecular hydride features taken from the
Kurucz (1995)\footnote{Available at http://kurucz.harvard.edu/}
database to form the initial line lists.
Then repeated syntheses of the solar spectrum (Delbouille, Neven, \& Roland
1973)\footnote{Available at http://mesola.obspm.fr/solar\_spect.php}
for $\lambda$~$>$~3000~\AA, and of the spectrum of the very metal-poor,
$n$-capture-deficient giant star HD~122563 (obtained as part of this
program), were used to refine the often poorly-known
transition probabilities of the atomic contaminants.
The line strengths of molecular hydrides (predominantly OH, less often
CH and NH) were varied together by changing the CNO abundances of the
individual stars until acceptable matches with the observed spectra
are obtained.

In this procedure, no alterations were permitted to the laboratory
data for the \ion{Ge}{1}, \ion{Os}{1}, \ion{Ir}{1}, and \ion{Pt}{1} lines.
Here we comment on the lines of each species employed in this survey
and the sources of their laboratory data.

{\it Germanium:} The best germanium abundance indicator is the \ion{Ge}{1}
3039.07~\AA\ line.
Following the discussion of Cowan \etal\ (2002), we adopted the
Bi\'emont \etal\ (1999) $gf$ value.
As Cowan \etal\ noted, two other strong \ion{Ge}{1} lines at 2651.17 and 
2651.57~\AA\ can be detected in the spectra of our program stars.
Unfortunately, they comprise parts of a large blended absorption feature
stretching over $\sim$2~\AA, and we could not derive reliable germanium
abundances from these lines.
Their line strengths do roughly correlate with that of the 3039~\AA\
line; probably they could become useful transitions with higher resolution
(R~$\simeq$~60,000) spectra.

{\it Zirconium:} Several \ion{Zr}{2} lines are detectable on our HST
spectra, allowing a link to be made with spectral features of this species
seen in the visible spectral region.
Cowan \etal\ (2002) and Sneden \etal\ (2003) have already demonstrated
that consistent abundances of Zr can be obtained from UV and
ground-based spectra.
Bi\'emont \etal\ (1981) published reliable $gf$-values for many
\ion{Zr}{2} lines, but unfortunately their study did not extend
to wavelengths below 3400~\AA.
Therefore, for the 3036.39+3036.51 \AA\ blend, and for the lines at 3054.84,
3060.12, and 3061.33 \AA, we adopted the transition probabilities
recommended by Kurucz (1998), who adjusted the original Corliss \& Bozman
(1962) values to conform with the B\'iemont \etal\ $gf$ scale.

{\it Osmium:} We analyzed the 2838.63 and 3058.18~\AA\ \ion{Os}{1} lines,
the same ones used in our original HST/GHRS study of osmium
in metal-poor giants (Sneden \etal\ 1998).
Spectra of these two lines are displayed in that paper's Figure~2.
We also added the line at 3301.57~\AA, which has been shown to be a
reliable \ion{Os}{1} abundance indicator in $r$-process-rich stars
(Cowan \etal\ 2002, Sneden \etal\ 2003).
Transition probabilities for the 3058 and 3301~\AA\ lines were adopted
from the recent laboratory analysis of Ivarsson \etal\ (2003):
log~$gf_{\rm 3058}$~=~--0.45, and log~$gf_{\rm 3301}$~=~--0.74.
These values are very close to those used in our previous studies
(--0.43, and --0.75, respectively; Kwiatkowski \etal\ 1984).
No recent laboratory study of the 2838~\AA\ \ion{Os}{1} line
exists, so we adopted the value used by Sneden \etal\ (1998):
log~$gf_{\rm 2838}$~=~+0.11, from Corliss \& Bozman (1962)
scaled to the Kwiatkowski \etal\ lifetime system.
None of the osmium lines are strong in the program star spectra,
obviating the need for hyperfine and isotopic substructure
calculations.

{\it Iridium:} We used the same \ion{Ir}{1} transitions as did Cowan \etal\
(2002):  3220.76, 3513.65, and 3800.12~\AA.
The original study employed $gf$ values based on the lifetime
measurements of Gough, Hannaford, \& Lowe (1983) and branching
ratios from Corliss \& Bozman (1962).
Happily, Ivarsson \etal\ (2003) also have provided transition
probabilities for two of the lines: log~$gf_{\rm 3513}$~=~--1.21, and
log~$gf_{\rm 3800}$~=~--1.44, which are in excellent agreement with
the older values of --1.26 and --1.45, respectively.
We adopted the Ivarsson \etal\ values for these lines (see further remarks
in the appendix), and  log~$gf_{\rm 3220}$~=~-0.52 for 
the third line (Gough et al. 1983).

{\it Platinum:} Den Hartog \etal\ (2005) have recently completed 
a new laboratory
transition probability analysis of \ion{Pt}{1}; their values are adopted here.
That study also searched for the cleanest and strongest lines in our
program star \bd17, originally studied in detail by Cowan \etal\ (2002).
This halo giant star was considered the most favorable case for \ion{Pt}{1}
detections, as it combines low metallicity ([Fe/H]~=~--2.1), large
$n$-capture/Fe abundance ratios (\eg, [Eu/Fe]~+0.9), relatively high
temperature (\teff~=~5200~K, which results in substantially weakened
molecular contaminating features), and excellent high-resolution
spectra in hand from 2500~\AA\ through 7000~\AA.
Of the 127 \ion{Pt}{1} lines with newly-determined $gf$ values, Den Hartog 
\etal\ found only 11 useful features for their platinum abundance study
of \bd17, and recommended only the relatively unblended lines at 2646.68,
2659.45, and 2929.79~\AA\ for application to other metal-poor stars.
Those lines were employed in the present study.

Hyperfine and isotopic splitting of \ion{Pt}{1} transitions must be taken
into account because the stronger lines can be on the damping
part of the curve-of-growth.
Platinum has six naturally occurring isotopes existing in solar-system
percentages of: \iso{192}{Pt}, 0.8\%; \iso{194}{Pt}, 32.9\%;
\iso{195}{Pt}, 33.8\%; \iso{196}{Pt}, 25.3\%; and \iso{198}{Pt}, 7.2\%.
The total isotopic splitting for some \ion{Pt}{1} lines can be as large
as $\simeq$0.03~\AA.
Additionally, the \iso{195}{Pt} lines split into 3--4 hyperfine structure
components that can have total separations of as much as $\simeq$0.09~\AA.
Wavelengths and $gf$ values of the resulting total of 8--9 transition
subcomponents have been taken from Table~7 of Den Hartog \etal\ (2005).
The solar isotopic mix has been adopted in all calculations.
In principle, $s$- and $r$-process $n$-capture nucleosynthesis could
produce different isotopic platinum mixes.
However, this element is a nearly pure $r$-process element
in solar-system material (95\%, in the most recent $r$/$s$ breakdown
of Simmerer \etal\ 2004).
Therefore, adoption of the solar-system platinum isotopic abundances
is appropriate for the generally $r$-process-rich stars of the present
sample.

{\it Lanthanum and Europium:} These two rare earth elements (Z~=~57 and 63,
respectively), often employed in investigations of the $r$- and $s$-processes
in the early Galaxy, were employed here to compare with the newly determined
very light and very heavy $n$-capture elements.
Six of our program stars were included in the Simmerer \etal\ (2004)
La and Eu survey, and their abundances are adopted here.
For stars HD~126587 and HD~175305 we have ground-based Keck~I HIRES spectra,
so we used synthetic spectrum computations, with the model atmospheres 
described above and the line lists of Simmerer \etal, to derive these
abundances from the \ion{La}{2} 3988.51, 3995.75, 4086.71, 4123.22, and
4333.75~\AA\ and the \ion{Eu}{2} 3819.67, 3907.11, 3971.97, 4129.72, and
4205.04~\AA\ transitions.
For BD+17$^{\rm o}$3248 and CS~22892-052, we adopted the La and Eu abundances 
of Cowan \etal\ (2002) and Sneden \etal\ (2003), respectively.
The only star unavailable to us for these elements was HD~6268.
We chose to employ the abundances of Burris \etal\ (2000) rather than the
more recent study of Honda \etal\ (2002) because the model atmosphere used
in our study has nearly the same parameters as those of Burris \etal

\subsection{Abundances of Ge, Zr, Os, and Pt}

Abundances were derived from synthetic/observed spectrum
matches for one line of \ion{Ge}{1}, four lines of \ion{Zr}{2},
and three lines each of \ion{Os}{1}, \ion{Ir}{1}, and \ion{Pt}{1}.
The abundances from individual lines are listed in Table~\ref{tab2}, and
the mean abundances are in Table~\ref{tab3}.
The low metallicity halo giant star HD~165195 (\teff~=~4235~K,
\logg~=~0.8, [Fe/H]~=~--2.4; Simmerer \etal\ 2004) was originally
observed as part of our HST/STIS program, but in the end it was
discarded.
Its temperature is about 300~K lower than any of the other stars,
producing a very strong-lined UV spectrum in spite of its low
metallicity.
No reliable abundances could be determined for this star.

Very few transitions have been used in the derivation of each
element's abundance.
All lines are at least partially blended in our program stars,
and many are weak features.
Therefore our mean abundance estimates given in
Table~\ref{tab3} have been computed according to the following rules.
Uncertain abundances from individual lines (designated with colons
in Table~\ref{tab2}) have been given equal weight with other abundances.
For a couple of stars where only upper limits could be determined
for some elements, the lowest values for the upper limits have been
quoted in Table~\ref{tab3}.
For \ion{Os}{1}, the 2838~\AA\ line is given half-weight; it may be
blended with an unidentified contaminant, since in the strongest-lined
stars it often yields a much larger abundance than do the 3058 and
3301~\AA\ lines.

Each of the three \ion{Ir}{1} lines has some blending issues that
can limit the iridium abundance reliability.
The 3800.12~\AA\ line is always very weak in the program stars, and
it lies in the large wing of the \ion{H}{1} 3797.9~\AA\ Balmer line.
The 3513.65~\AA\ line is stronger, but unfortunately it is sandwiched
in between the very strong 3513.49~\AA\ \ion{Co}{1} and 3513.83~\AA\
\ion{Fe}{1} lines; see Figure~1 (lower panel) of Sneden \etal\ (2000)
for the appearance of this feature in CS~22892-052.
However, iridium abundances derived from syntheses of these two lines
are well-correlated in all program stars.
The 3220.76~\AA\ line should be the intrinsically strongest transition.
However, attempts to synthesize this feature with inclusion of only
\ion{Ir}{1} absorption yielded iridium abundances that were 0.3-0.4~dex
larger than the means of the abundances derived from the 3513 and
3800~\AA\ lines.
Additionally, the observed feature appears to be significantly contaminated
by another absorber at approximately 3220.72~\AA, which could be an
\ion{Fe}{1} line listed in the Kurucz (1998) database.
We arbitrarily set the $gf$-value of this line to a value that provided
a reasonable total fit to the overall feature in typical program stars,
but the derived iridium abundance uncertainties here remain large, so
the 3220~\AA\ line abundance is entered at half-weight into the final
averages.

The \ion{Pt}{1} 2646~\AA\ line also appears to be significantly blended
in the stronger-lined stars, compared to the 2659 and 2909~\AA\ lines.
It too is entered into the means with half-weight.

\subsection{Uncertainties}

In much of our previous work on $n$-capture-rich stars we have concentrated
on the rare-earth elements, 57~$\leq$~Z~$\leq$~72.
Nearly all detectable transitions of these elements in cool stars arise from 
low excitation levels of the first ions.
Since the ionization potentials of the rare earths are relatively low 
(5.5--6.8~eV), these elements exist almost exclusively as the first ions.
In LTE the Boltzmann and Saha factors are essentially the same
for all rare-earth transitions, leading to almost zero sensitivity to 
\teff\ and \logg\ in the element-to-element abundance ratios (\eg, see
Table~3 of Westin \etal\ 2000).

Most of the species of the present study arise from low excitation
     levels of the neutral atoms.
     Therefore, abundance inter-comparisons among Ge, Os, Ir, and Pt
     also have little sensitivity to model atmosphere parameters.
     Linking these abundances to those of the rare earths, however, involves a
     neutral versus first ion comparison.
     The average first ionization potential,
     $<$I.P.(Ge,Os,Ir,Pt)$>$~$\simeq$8.5~eV, is significantly larger than
     the values for most Fe-peak elements.
     Thus these elements exist substantially as \ion{Ge}{1}, \ion{Os}{1},
     \ion{Ir}{1}, and \ion{Pt}{1}.
     Their responses to changes in atmospheric parameters are different
     than those of Fe-peak neutral species.
Repeated computations with variations in \teff, \logg, [M/H], and \vt\ for 
these elements as well as for typical \ion{Fe}{1} and \ion{Eu}{2} lines
yielded the following average abundance variations (rounding to the 
nearest 0.05~dex) in the atmospheric parameter domain of our program stars.
For $\delta$\teff~=~$\pm$150~K: 
$\delta$[Fe/H]~$\simeq$~$\pm$0.15, 
$\delta$[Eu/H]~$\simeq$~$\pm$0.10,
$\delta$[Ge,Os,Ir,Pt/H]~$\simeq$~$\pm$0.20, or 
$\delta$[Eu/Fe]~$\simeq$~$\mp$0.05 and
$\delta$[Ge,Os,Ir,Pt/Fe]~$\simeq$~$\pm$0.05.
For $\delta$ \logg~=~$\pm$0.3: 
$\delta$[Fe/H]~$\simeq$~$\mp$0.15,
$\delta$[Eu/H]~$\simeq$~$\pm$0.05,
$\delta$[Ge,Os,Ir,Pt/H]~$\simeq$~$\pm$0.00, 
or $\delta$[Eu/Fe]~$\simeq$~$\mp$0.10 and
$\delta$[Ge,Os,Ir,Pt/Fe]~$\simeq$~$\pm$0.15.
There is negligible dependence on metallicity through variations of this
model parameter by $\delta$[M/H]~$\simeq$~$\pm$0.30.
Finally, the $n$-capture transitions under consideration here are relatively
weak, limiting microturbulence dependence to $<$0.1~dex for 
$\delta$\vt~=~$\pm$0.2~\kmsec.

Given these dependencies, the total influence of atmospheric parameter
uncertainties in the abundance ratios of Ge, Os, Ir, and Pt to Fe or Eu
is $\simeq\pm$0.20.
Adding in typical uncertainties of fit in matching synthetic and observed
spectra of $\simeq$0.15 yields total suggested abundance uncertainties
of $\simeq$0.25.
However, abundance ratios taken simply among Ge, Os, Ir, and Pt should be more
reliable, since they all arise from low-excitation states of the neutral 
species.  
In particular, the very low relative abundances of Ge to the Os-Ir-Pt group
or to Fe cannot be explained away from atmospheric parameter uncertainties.

The Zr abundances have been derived from low-excitation ionized-species 
transitions, and thus can be compared directly to Eu abundances with 
little concern about model atmosphere uncertainties.  
This element has been studied in many of our stars using \ion{Zr}{2} lines
at longer wavelengths.
In Table~\ref{tab4} we summarize the comparisons between our HST-based Zr
abundances and other results.
For the present study, Zr abundances have been derived from synthetic/observed
spectrum matches of the \ion{Zr}{2} lines at 3998.96, 4050.33, 4090.51, 
4208.98, and 4496.97~\AA\ lines appearing our Keck HIRES-I data.
For nine stars with both UV and visible-wavelength spectra, we derive
$<\delta$[Zr/Fe]$>$~= --0.08~$\pm$~0.06 ($\sigma$~=~0.17), where 
$\delta$[Zr/Fe] is in the sense HST/STIS $minus$ Keck~I HIRES.
The survey of Burris \etal\ (2000) has eight stars in common with the present
work, for which we derive $<\delta$[Zr/Fe]$>$~= --0.08~$\pm$~0.06 
($\sigma$~=~0.17), in the sense this study $minus$ Burris \etal.
Some individual stars  have been subjected to very detailed analyses. 
Inspection of Table~\ref{tab4} suggests that the HST/STIS Zr abundances
are systematically lower than most visible-wavelength literature values
by $\sim$0.1~dex.
This offset cannot be pursued further here, because modern lab $gf$ studies
of the $UV$ lines of \ion{Zr}{2} have yet to be accomplished.

\section{Discussion}

Our new observations -- HST/STIS spectra of the elements Ge, Zr, Os 
and Pt, and Keck~I HIRES spectra of Ir -- allow us to make $n$-capture 
abundance comparisons among the sample of metal-poor Galactic halo stars.
We have also incorporated these new detections with previously 
determined values for other elements to obtain detailed $n$-capture 
elemental abundance distributions for the well-studied and $r$-process 
rich stars \cs22\ and \bd17.  

\begin{figure}
\epsscale{0.90}
\plotone{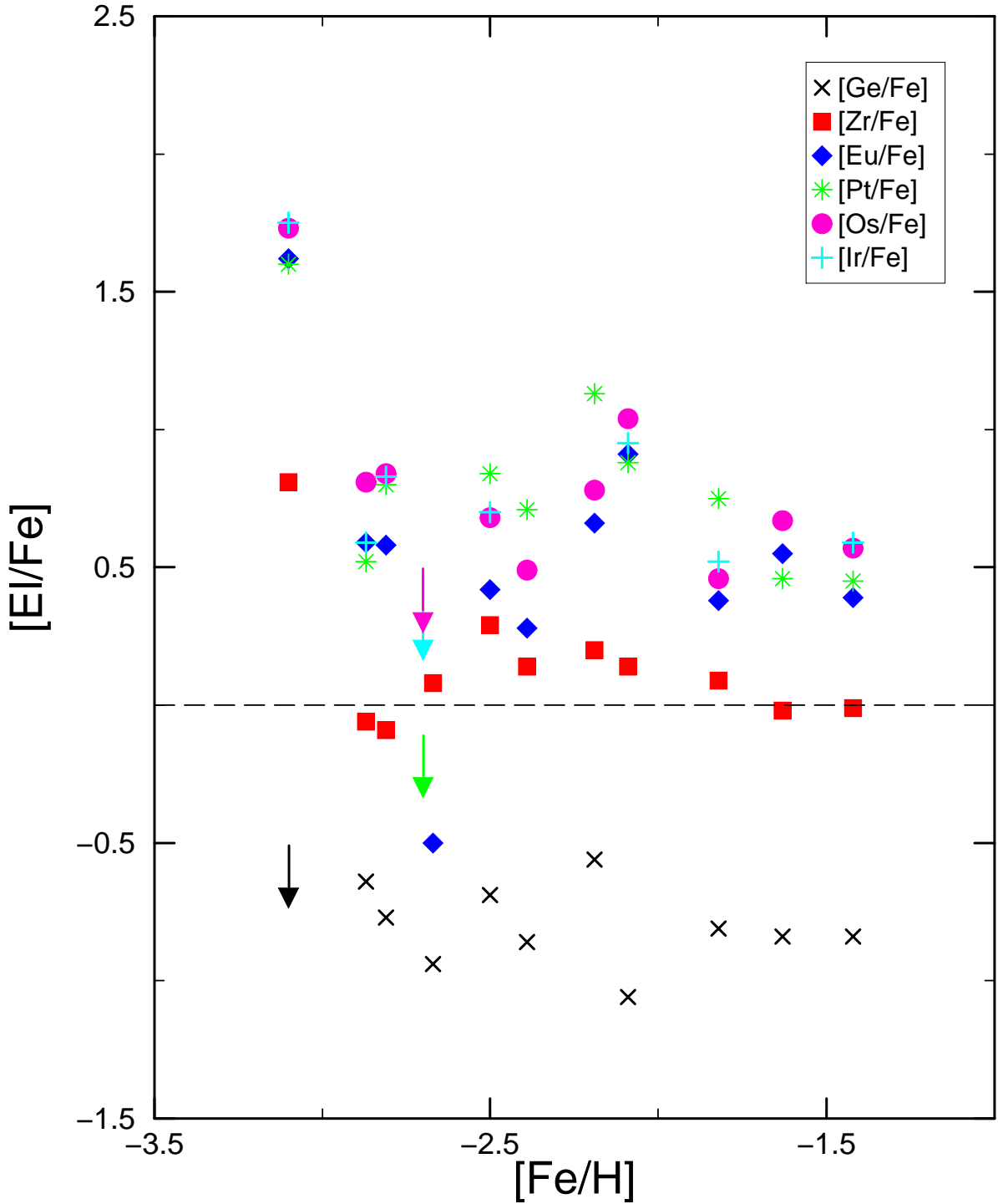}
\vskip -0.8truein
\caption{All of the $n$-capture abundances (or upper limits) determined 
in this study, plotted as a function of [Fe/H] metallicities.
for the $n$-capture elements Ge, Zr and Eu in a group of halo stars.
The symbols are defined in the figure legend, and the horizontal dashed
line represents the solar abundance ratio, [El/Fe]~=~0.0.
Of the four upper limits (downward-pointing arrows) 
displayed in this figure, three of them are for
the 3$^{\rm rd}$ $n$-capture-peak elements Os, Ir, and Pt in the 
$r$-process-poor star HD~122563.
The other upper limit is for Ge in the extreme $r$-process-rich star 
CS~22892-052.
\label{fig1}}
\end{figure}

\begin{figure}
\epsscale{0.95}
\plotone{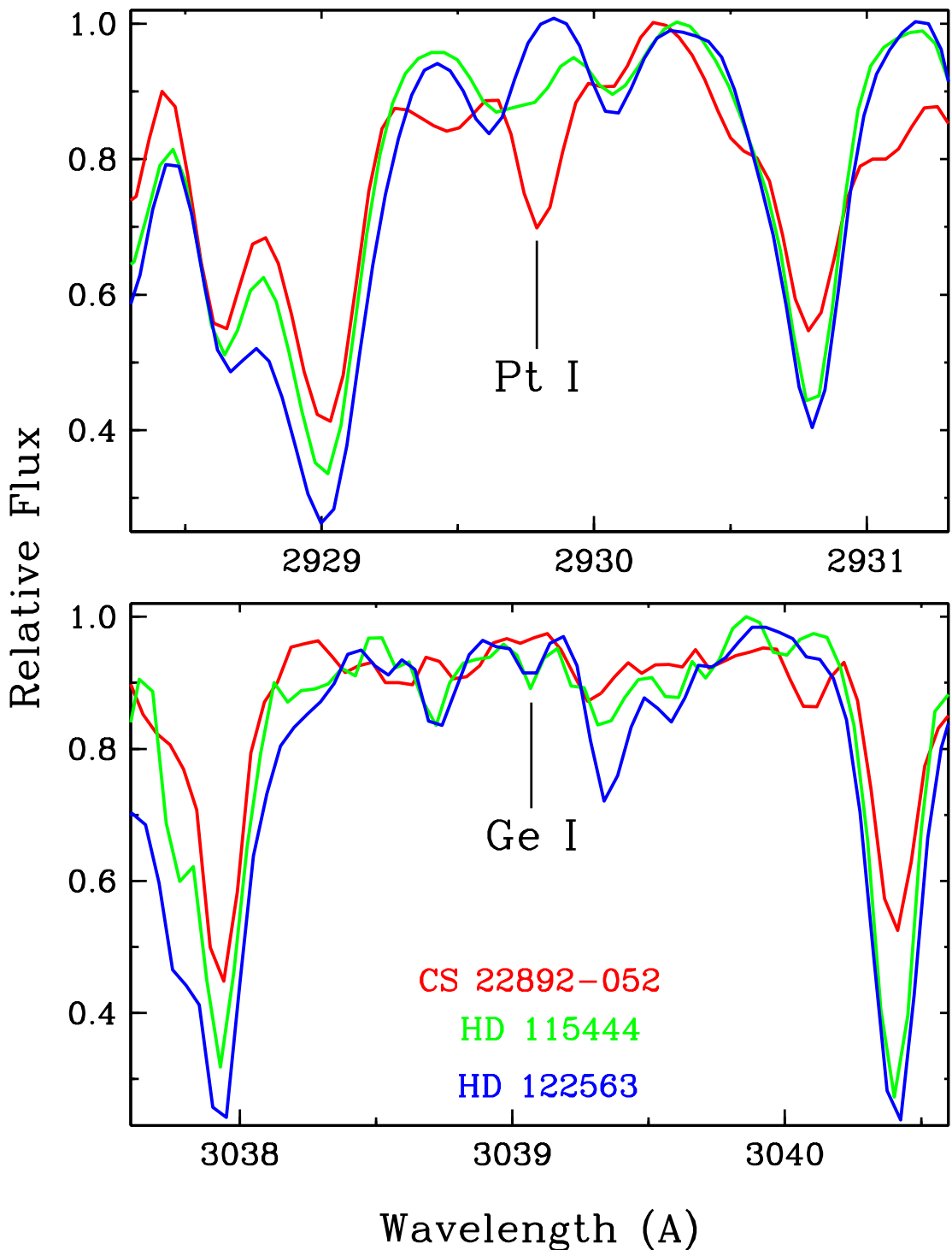}
\vskip -.5truein
\caption{Small regions of HST/STIS spectra centered on the 2929.80~\AA\ 
Pt~I line (top panel) and the 3039~\AA\ Ge~I line (bottom panel), 
in three similar-metallicity program stars: HD~122563 ($r$-process-poor), 
HD~115444 ($r$-process-enhanced), and CS~22892-052 
(extremely $r$-process-rich).
The Pt~I line strengths agree well with the general progression 
of relative $r$-process abundance levels; 
The Ge~I line strengths are essentially uncorrelated with them.
\label{fig2}}
\end{figure}

In Figure~\ref{fig1} the entire abundance set is summarized by plotting
relative abundance ratios [El/Fe] versus [Fe/H] metallicities.
A rough progression of increasing abundance ratio with increasing atomic
number is evident: large deficiencies of the light $n$-capture
element Ge in all program stars; weak or no enhancements of the 
intermediate-mass element Zr in all stars except the extreme $r$-process-rich
star \cs22; and large overabundances of the heaviest elements Os, 
Ir, and Pt in all stars except the $r$-process-poor star HD~122563.
A similar conclusion may be seen in Figure~\ref{fig2}, in which we show
the observed spectra of three stars with contrasting $r$-process abundance 
levels: HD~122563 ([\ion{Fe}{1}/H]~=~--2.61, [Eu/Fe]~=~--0.50; 
see Tables~\ref{tab1} and \ref{tab3}, and their literature references); 
HD~115444 (([\ion{Fe}{1}/H]~=~--2.71, [Eu/Fe]~=~+0.58); and
\cs22\ (([\ion{Fe}{1}/H]~=~--3.09, [Eu/Fe]~=~+1.62).
While the relative absorption strengths of the \ion{Pt}{1} lines of these
three stars qualitatively track the [Eu/Fe] values (top panel), no such 
correlation is seen in the \ion{Ge}{1} line strengths (bottom panel).
In fact, \cs22\ has the largest [Eu/Fe] value but clearly the weakest
\ion{Ge}{1} feature.
In the following subsections we amplify and interpret these observational
results.

\subsection{Abundances of the Heaviest $n$-Capture Elements}
                                                                                
Earlier work (Cowan \etal\ 1996; Sneden \etal\ 1998, 2003) used HST with 
STIS and the Goddard High Resolution Spectrograph to detect \third\ 
$n$-capture-peak ({\it e.g.}, Pt) in  a few individual metal-poor halo stars.
Our new abundance data for Os, Ir, and Pt allow us to make the first 
moderate-sample systematic study of the heaviest stable $n$-capture elements, 
and to compare them with the $n$-capture element Eu, which is synthesized 
almost entirely in the $r$-process.

\begin{figure}
\epsscale{0.95}
\plotone{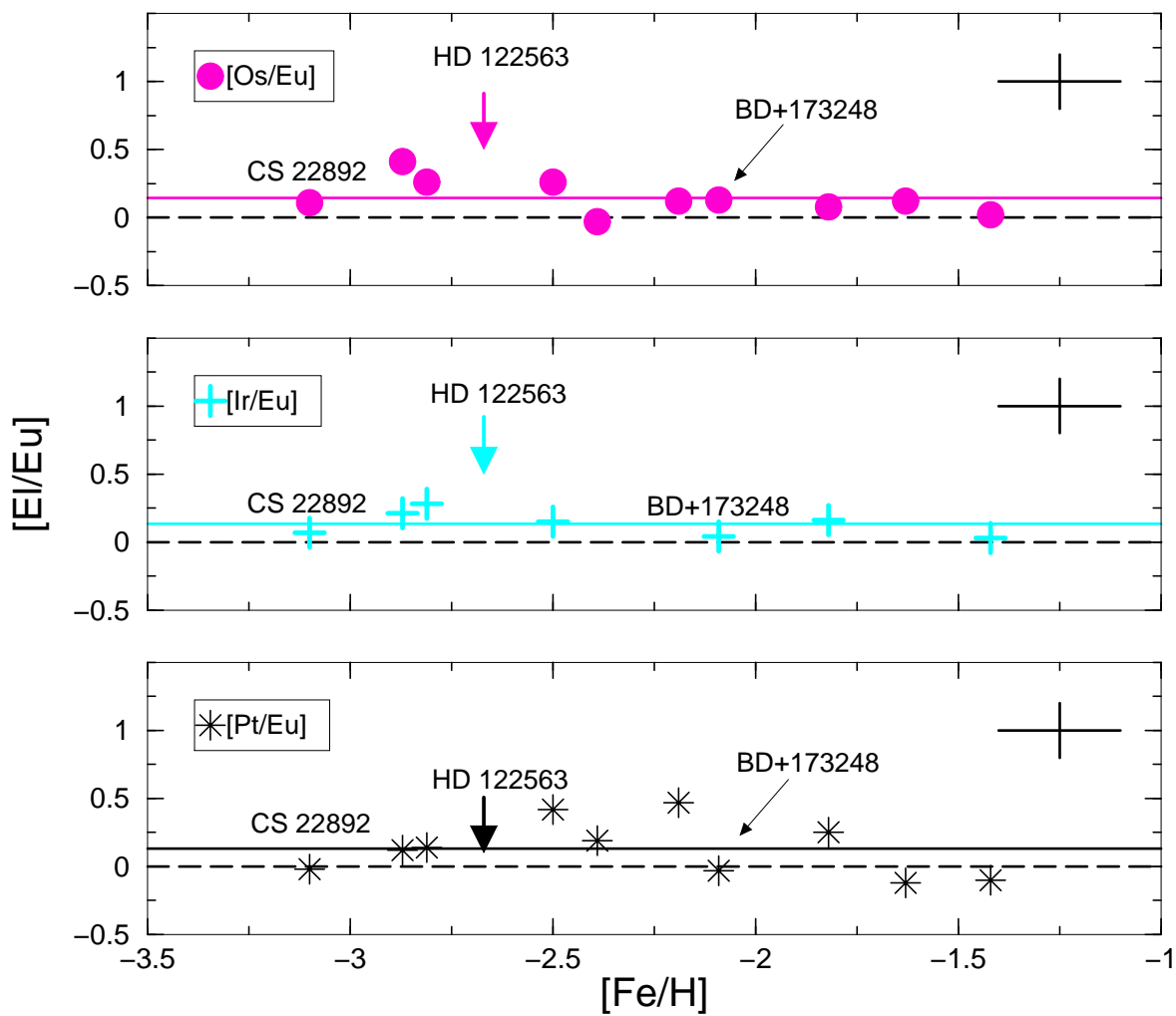}
\caption{Relative abundance ratios [Os/Fe] (top panel), [Ir/Fe]
(middle panel), and [Pt/Fe] (bottom panel) compared to the standard 
$r$-process abundance indicator [Eu/Fe].
Dashed lines are drawn in each panel at [El/Fe]~=~0.0 to indicate the solar 
abundance ratio, and the solid lines shows the mean derived values, 
$<$[El/Fe]$>$. A typical error is indicated by the cross in each panel.  
\label{fig3}}
\end{figure}

We make direct comparisons of Os, Ir, and Pt abundances with Eu in
the three panels of Figure~\ref{fig3}.
While the ratios [El/Fe] shown in Figure~\ref{fig1} all indicate substantial
overabundances of these three elements, the comparisons to Eu in 
Figure~\ref{fig3} demonstrate the clearly correlated abundance behavior 
of Eu, Os, Ir, and Pt.
The very small deviations from [El/Eu]~=~0.0 indicated by the solid horizontal
lines of each panel 
($<$[Os/Eu]$>$~=~+0.15, $\sigma$~=~0.12; 
 $<$[Ir/Eu]$>$~=~+0.13, $\sigma$~=~0.09;, and 
 $<$[Pt/Eu]$>$~=~+0.13, $\sigma$~=~0.20)
suggest that the solar-system $r$-process abundance distribution is mimicked
in our sample (and possibly all) $r$-process-rich stars born in the early
Galactic halo.
We regard the mean $\sim$+0.15~dex offset as observationally 
indistinguishable from [El/Eu]~=~0.0; see the uncertainty discussion of \S3.3.
These very-heavy-element abundance comparisons strongly suggest a similar 
synthesis origin for Eu, Os, Ir, and Pt in the $r$-process sites that were 
the progenitors to the observed halo stars.
                                                                                
We also want to note the La/Eu ratios listed in 
Table~\ref{tab3} (see also the discussion in \S 3.2). 
In solar system material La (dominantly an $s$-process element, 
see Simmerer \etal \ 2004) is more abundant than the $r$-process element
Eu. As is seen in the  data compiled in  Table~\ref{tab3}, however,   
$<$[La/Eu]$>$ =  --0.4, clearly indicating
that all of the stars in our sample are $r$-process-rich.
(See Simmerer \etal \ for further discussion of the synthesis of La 
in these stars.)

\subsection{The Light $n$-Capture Elementals Germanium and Zirconium}

\begin{figure}
\plotone{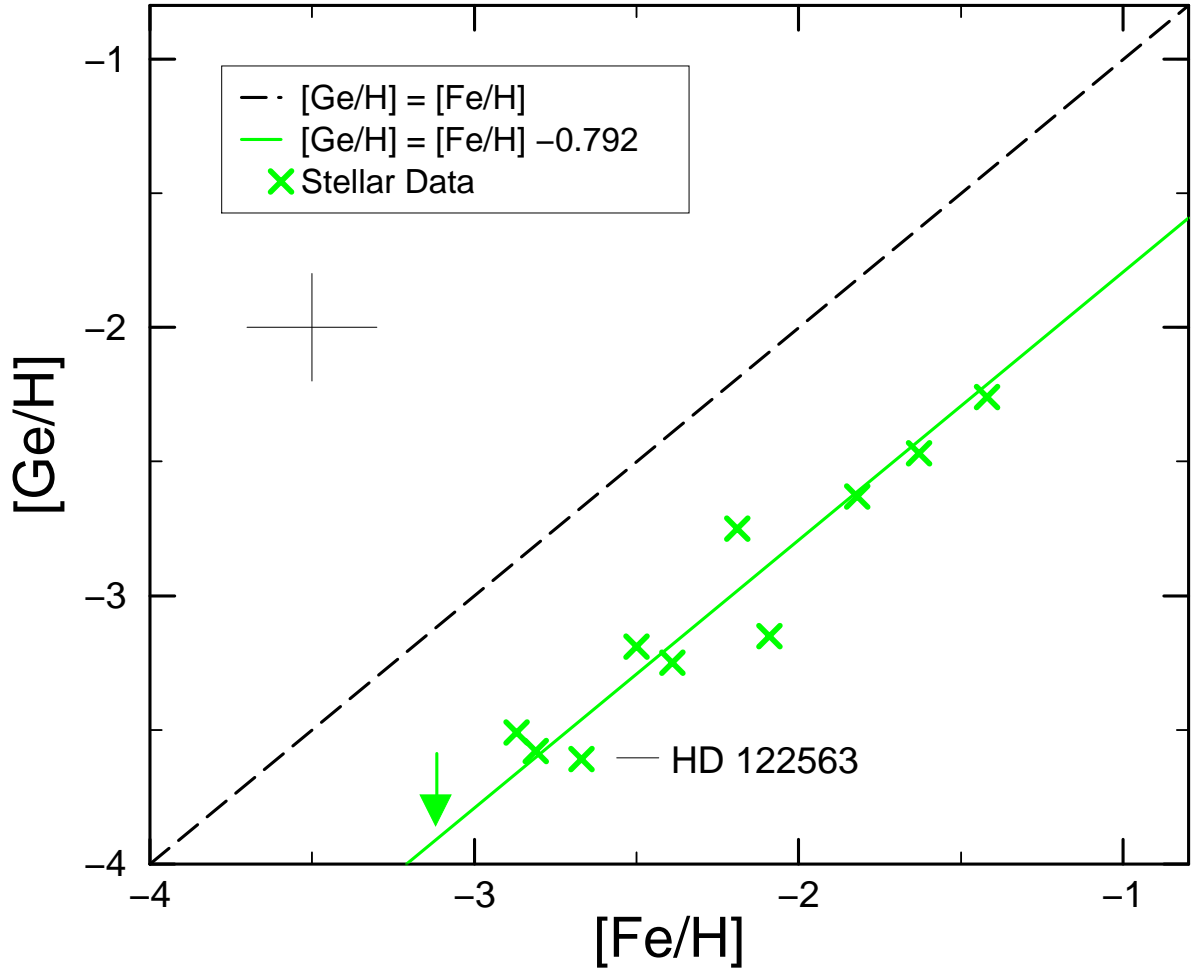}
\caption{Relative abundances [Ge/H] displayed as a function of [Fe/H] 
metallicity for our sample of 11 Galactic halo stars. 
The arrow represents the derived upper limit for \cs22. 
The dashed  line indicates the solar abundance ratio of these elements,
[Ge/H]~=~[Fe/H], while the solid green line shows the derived correlation 
[Ge/H]~= [Fe/H]~--~0.79.
A typical error is indicated by the cross.
\label{fig4}}
\end{figure}

\begin{figure}
\plotone{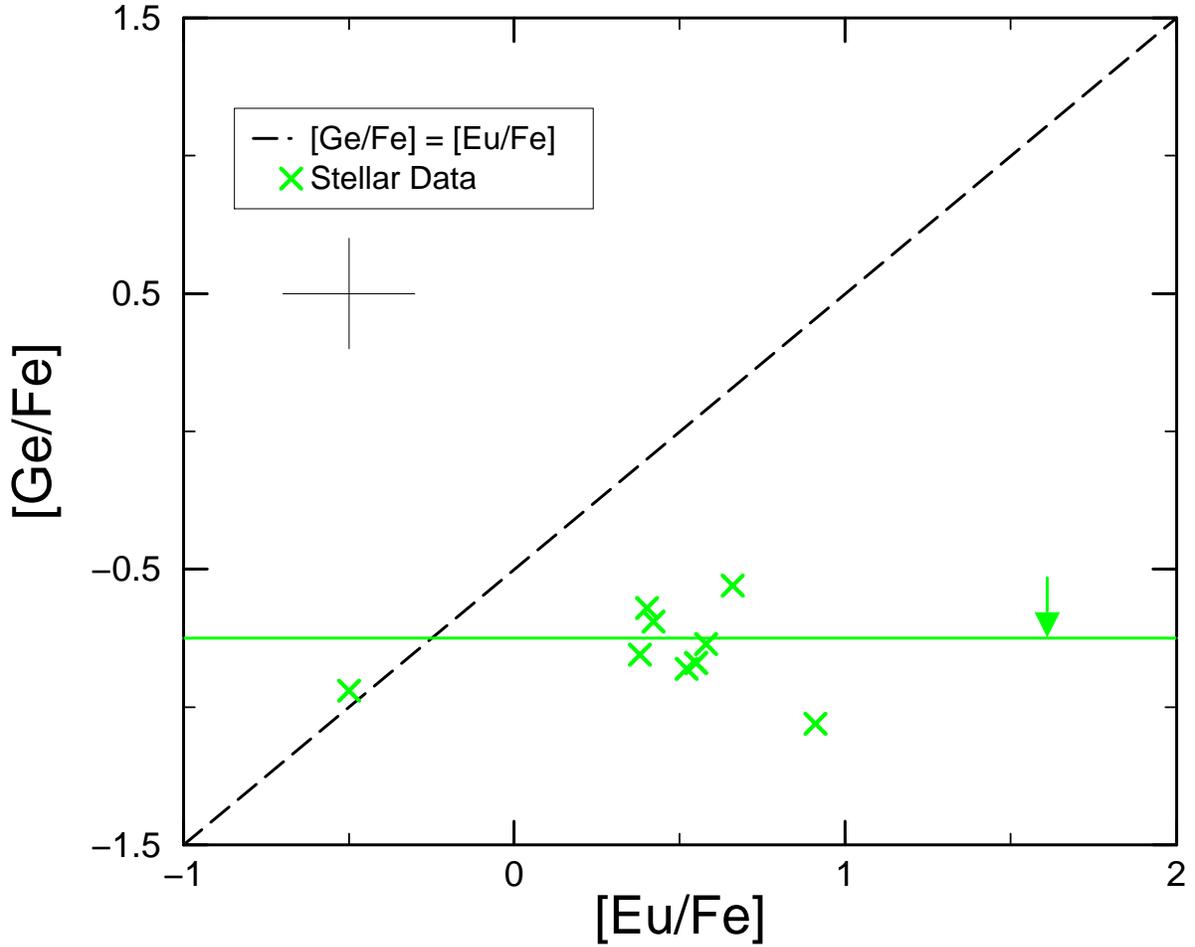}
\caption{Correlation between the abundance ratios [Ge/Fe] and [Eu/Fe].
The dashed line indicates a direct correlation
between Ge and Eu abundances. 
As in the previous figure, the arrow represents the derived upper 
limit for \cs22.  
The solid green  line at [Ge/Fe]r~--~0.79 is a fit to the observed data.
A typical error is indicated by the cross.
\label{fig5}}
\end{figure}

In Figure \ref{fig4} we plot [Ge/H] values with respect to the traditional
metallicity indicators [Fe/H]. 
It is easy to see that the Ge abundances scale with metallicity but at a 
depressed (with respect to solar) level:
$<$[Ge/H]$>$~=~[Fe/H]~--~0.79$\pm$0.04 ($\sigma$~=~0.14). 
Further supporting this interpretation are abundance comparisons of 
Ge with respect to the $n$-capture element Eu that we illustrate in 
Figure~\ref{fig5}.
If Ge and Eu were correlated, the abundances would fall along the
straight (diagonal) line illustrated in the figure. 
Obviously the abundances of Ge seem to be uncorrelated with those of the 
$r$-process element Eu.  
In fact, [Ge/Fe] for the $r$-process poor star HD~122563 is comparable 
to the values found for $r$-process ({\it i.e.}, Eu) rich stars, including 
the upper limit for \cs22.  
While $n$-capture processes are important for Ge production in solar system
material (\eg, Simmerer \etal\ 2004), these abundance comparisons 
immediately suggest a different origin for this element early in the 
history of the Galaxy.  

Our abundance data appear to be more consistent with an explosive (or 
charged-particle) synthesis for Ge.
This might occur as a result of capture on iron-peak nuclei, perhaps 
during the so-called ``$\alpha$-rich freeze-out'' in a supernova 
environment. 
However, 
calculations to date (Nakamura \etal\  1999; Hoffman \etal\  2001; Heger \&
Woosley 2002; Umeda \& Nomoto 2005; Chieffi \& Limongi 2004) have difficulties
explaining the observed trends in [Ge/Fe]. Indeed, the trends in iron-peak
nuclei revealed in Cayrel \etal\ 2004) strongly
suggest the occurrence
of a rather dramatic alpha-rich freeze associated with the earliest and most
metal-deficient stellar populations, down to at least a metallicity level
[Fe/H] $\sim$ -4 (Truran \etal\  2005). The strong 
temperature (density) dependence
of this process indicates that fine tuning may be required to fit the
observations. This may
explain the fact that existing studies have not reproduced the trends.
It is interesting to note, in this regard, that the trends seen in the
two "hyper"-iron-poor stars (Frebel \etal\ 2004; Christlieb \etal\ 2004)
do not appear to be consistent with the extrapolation of those reflected
in the Cayrel \etal\ (2004) study below [Fe/H] $\sim$ -4.
This contribution to Ge synthesis
appears to be the dominant production mechanism at low 
metallicities.
Of course, un-tracebly small contributions from $s$- and/or $r$-process 
production cannot be ruled out (but significant $s$-process production 
is not expected at [Fe/H]  $\ltaprx$ --2).
At higher metallicities and the onset of the $s$-process,
it would be expected that the Ge production would increase sharply
and no longer be correlated with the iron production (Gallino, private
communication).

\begin{figure}
\plotone{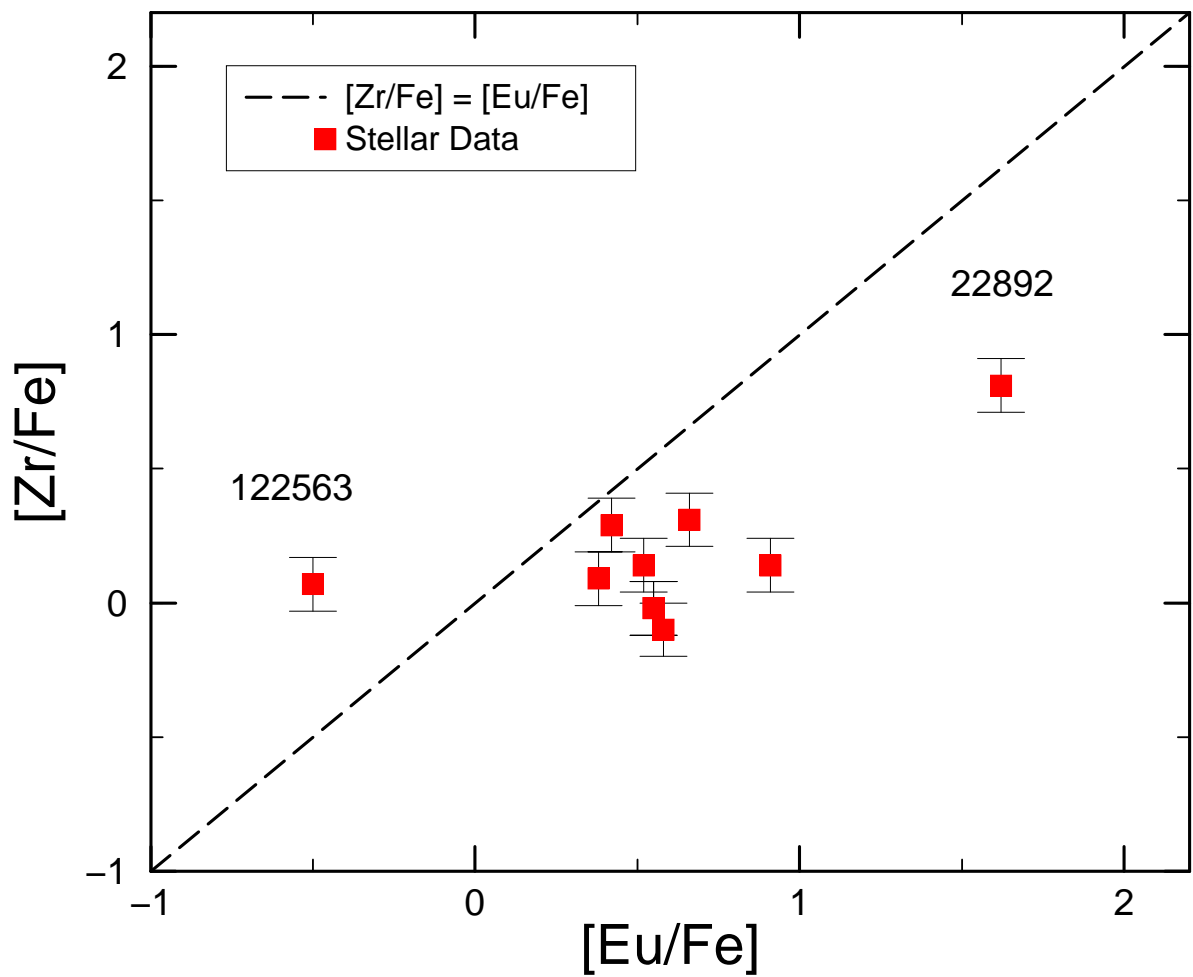}
\caption{Correlation between the abundance ratios of [Zr/Fe] (obtained
exclusively with HST/STIS) and [Eu/Fe].
The dashed  line indicates a direct correlation between Zr and 
Eu abundances. 
\label{fig6}}
\end{figure}

We make a similar abundance comparison of Zr versus Eu for our sample 
of stars in Figure \ref{fig6}. 
As discussed in \S3.3, we note that the Zr abundances obtained with HST/STIS 
are well correlated with the ground-based abundances of this element.
The abundance data, excluding \cs22, show very little scatter and appear to 
be uncorrelated with Eu -- this includes stars with [Eu/Fe]~$\simeq$~1.0. 
The exception is \cs22, for which [Zr/Fe] is substantially higher than
in, for example, HD~122563. 
The Zr and Eu abundances, however, do not follow a clearly linear correlation 
in the figure.
This again would seem to indicate a different synthesis origin for these 
two elements, something recently discussed in more detail by Travaglio 
\etal\ (2004). 
Their more extensive abundance analysis suggests that known $n$-capture 
processes can explain some of the production of Zr, but that the 
nucleosynthetic origin of this element is different than that for heavier
$n$-capture elements such as Ba and Eu. 
Furthermore, Travaglio \etal\ argue that an additional (lighter element) 
primary process is also responsible for some fraction of the synthesis of 
Sr and Y.
We note finally that in addition to our abundance determination for 
\cs22, the halo giant CS~31081--001 (the Hill \etal\ 2002 ``uranium'' star) 
also has very enhanced [Eu/Fe] $and$ [Zr/Fe]. 
It may be that in these very $r$-process-enhanced stars, this kind of
$n$-capture nucleosynthesis contribution overwhelms the light primary
process proposed by Travaglio \etal\ -- perhaps as a result 
of fission recycling. We note that 
the relative constancy of [Zr/Fe] with both [Fe/H] and [Eu/Fe] supports the 
conclusion of Johnson and Bolte (2002) - based on the constancy of the ratio
[Y/Zr] with respect to [Zr/Fe] and [Ba/Fe] - that the source of Zr in 
metal-poor stars must be the same for both the $r$-process-rich and 
$r$-process-poor stars.

\subsection{Heavy Element Abundance Scatter}

The heavy element abundance patterns presented here exhibit striking 
differences as a function of metallicity.
The linear correlation (with very small scatter) of Ge and Fe abundances 
suggests that Ge must have been produced rather commonly in stars -- 
even at early times in the Galaxy -- over a wide range of metallicity.
The Zr abundances show much the same behavior as Ge with (perhaps) 
somewhat more scatter, suggesting some variations in abundance with respect 
to Fe.
The pattern of the heavy $n$-capture elements Eu, Os, Ir and Pt is however
very different than that of Ge and Zr. 
There is a very large star-to-star scatter in the abundance values with 
respect to iron, particularly at low metallicity -- a factor of $>$ 100  
at [Fe/H]~$<$~--2.5. (We note that while there is only one star, 
HD 122563, in our sample with a very low [Eu/Fe] ratio, observations
(see {\it e.g.}, Burris et al. 2000) have indicated other such stars.
At higher metallicities this scatter diminishes dramatically.
This general notion is of course not new, having been seen previously for Eu 
(see e.g., Gilroy \etal\ 1988; Burris \etal\ 2000; Sneden \& Cowan 2003). 
This is the first clear indication that \third\ $r$-process peak 
elements also show the same scatter. 

These apparently conflicting trends can be explained by assuming that 
at early times (and some low metallicities) the Galaxy was
chemically inhomogeneous with some regions containing larger amounts of
$r$-process ejecta than others. 
Then at higher metallicities (and later times) these differences in the 
total abundance levels would be minimized -- this would be as a result 
of a higher number of events, which would produce an abundance average,  
and probably mixing throughout the Galaxy.
Thus, from Ge to Zr to Eu-Os-Ir-Pt we might be witnessing decreasing event
statistics, \ie, a smaller number of (supernova) sites at very
low metallicities, which create these elements (Cowan \& Thielemann 2004
and references therein).
This might further indicate that not all supernovae, or at least those 
that make lighter $n$-capture elements like Ge and Zr, are responsible for 
synthesizing the heavier $n$-capture elements (these $r$-process events 
would have been rare) at very low metallicities early in the history of 
the Galaxy. 
Such abundance comparisons and scatter (\eg, [Eu/Fe] versus [Fe/H]) are 
also providing  new  clues into the earliest stars and the chemical 
evolution of the Galaxy (\eg, Wasserburg \& Qian 2000; Fields, Truran 
\& Cowan 2002) and the nature of and site for the $r$-process, particularly 
early in the Galaxy (Argast \etal\ 2004).

\subsection{Elemental Abundance Distributions in Individual Stars}

\begin{figure}
\plotone{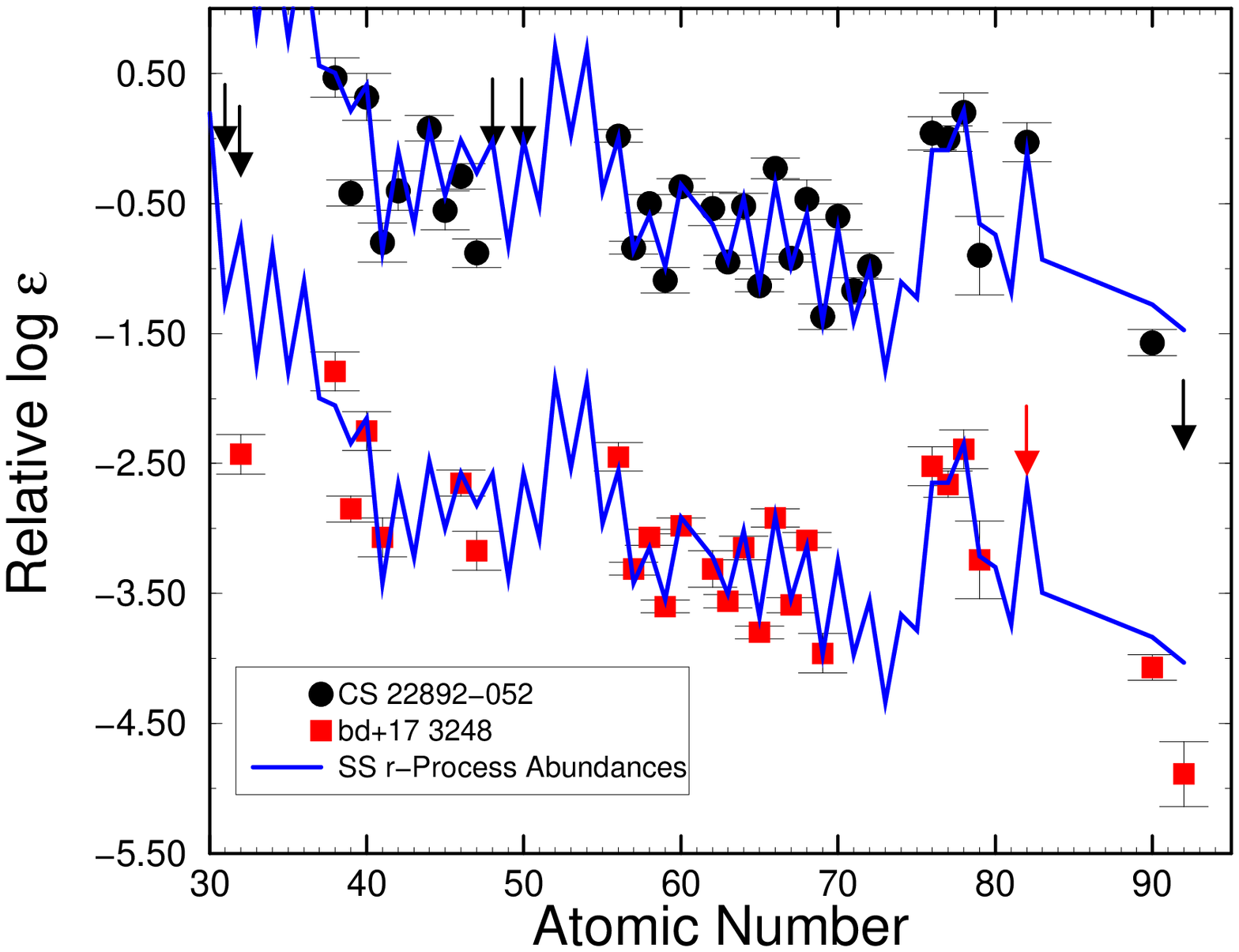}
\caption{The neutron-capture elemental abundance
pattern
in the Galactic halo stars CS 22892--052 and \bd17 compared with
the (scaled) solar system {\it r}-process abundances (solid line).
The abundances of \bd17 have been vertically scaled downward for 
display purposes. 
\label{fig7}}
\end{figure}

Two of our sample stars are the very $r$-process rich stars \cs22 and \bd17.
Employing our new observations and new, more reliable abundance 
determinations for Nd (Den Hartog \etal\ 2003) and Ho (Lawler, Sneden,
\& Cowan 2004),  
we have updated and supplemented the (ground-based and HST)
abundances previously obtained for \cs22\ (Sneden \etal\ 2003) and 
\bd17 (Cowan \etal\ 2002).
We show in Figure~\ref{fig7} the detailed abundance distributions 
for both of these stars. 
The values for \bd17 have been arbitrarily displaced downward for display 
purposes. 
We also show for comparison the solar system $r$-process abundances
(solid lines), determined based upon the classical $s$-process model and 
utilizing the most recent $r$-/$s$-process deconvolution reported by 
Simmerer \etal\ (2004).  
Several points are worth noting in this figure.  
The agreement between the rare earth elements (\eg, Ba and Eu) and the 
solar system $r$-process abundances is now seen to extend into, and includes, 
the \third \ $r$-process peak elements Os--Pt in \cs22 and \bd17. 
Note that employing the new atomic experimental data of Den Hartog \etal\ (2005) 
results in a shift downward of 0.1 dex -- with respect to the previously
determined value --  in the abundance of Pt. 
This element (along with Os and Ir) now falls on the same scaled solar 
system $r$-process curve that also matches the abundances of the 
rare-earth elements such as Eu, strengthening the apparent synthesis
connection between these heavier $n$-capture elements.

Our new, more reliable abundance determinations for the elements Nd and Ho 
are also consistent with the solar system $r$-process distribution.  
However, as shown in Figure~\ref{fig7}, the abundances of elements 
with Z~$<$~56 (\ie, below Ba) in general fall below the scaled solar 
$r$-process curve for \cs22. 
There is very little data available for \bd17\ in this region of 
40~$\leq$~Z~$\leq$~50, but its Ag abundance, in particular, 
is much less than the scaled solar-system $r$-process curve.
Only upper limits on Ge and Ga were obtained with HST for \cs22, but those 
abundances fall far below the solar curve, as does the Ge abundance in \bd17.
This indicates that there may be two processes and may suggest two 
astrophysical sites for $r$-process nucleosynthesis -- with one for 
lighter and another for the heavier $n$-capture elements.
This possibility was suggested earlier (see Wasserburg, Busso, \& Gallino 
1996 and Wasserburg \& Qian 2000) with supernovae with different masses 
and frequencies responsible for the two ends of the $n$-capture abundance 
distribution.
Other models have suggested neutron-star binary mergers as one of the 
possible sites, particularly for the heavier $n$-capture elements where 
supernovae models have had difficulties in achieving  the required high 
entropies (Freiburghaus \etal\ 1999; Rosswog, Rosswog, \& Thielemann
1999; but see Argast 
\etal\ 2004).
In addition to a combination of supernovae and neutron-star mergers,
it has been suggested that the light and heavy $n$-capture elements 
could also be synthesized in the same core-collapse supernova 
(Cameron 2001; see also recent reviews by Cowan \& Sneden 2004 and 
Cowan \& Thielemann 2004 for further discussion).

{\section{Conclusions}}

We have made new detections of the elements Ge, Zr, Os, and Pt
with HST (STIS), along
with Ir using Keck I (HIRES), in a sample of metal-poor Galactic halo
stars. These are the first large-sample abundance determinations of
these elements in such stars.
The abundances of the elements Os, Ir and Pt, in the \third $r$-process
peak  appear to be correlated 
among themselves and with  Eu -- the extensively observed,  
rare-earth, $r$-process element -- 
indicating  a similar nucleosynthesis origin and site. 
In contrast the  
Ge abundance  appears to scale with iron in the halo stars and is
independent of the Eu abundances in those stars. This suggests an
explosive,  rather than $n$-capture, synthesis for this element
in  stars at very low
metallicities. Perhaps this might be the result of some type of
alpha-rich freeze out in SNe early in the history of the Galaxy -- 
at higher ({\it e.g.}, near solar) 
metallicities, and later times, it would be expected that  
$r$- and  $s$-process production would dominate production.
The Zr abundances in the sample stars 
do not in general scale with metallicity, nor with 
the Eu abundances, suggesting a different origin for this element than
for the heavier $n$-capture elements. 
The one exception is for the very $r$-process rich 
(high Eu abundance) star \cs22,  where there is a significant increase
in the Zr abundance.  This element, however, has a complicated synthesis
with perhaps several processes contributing 
(see {\it e.g.}, Travaglio et al. 2004).

The star-to-star abundance scatter  
among the elements is quite different, 
with the lighter element Ge  showing  very little scatter over the range
of metallicity studied. This suggests a common origin - most SNe making this
element - even at very low metallicities.
Zr like Ge also shows little scatter with iron except for the case again
of \cs22. 
It has been shown  previously that [Eu/Fe] exhibits a large star-to-star
scatter as a function of metallicity (see {\it e.g.} Burris et al. 2000). 
Our results are the first demonstration that the 
abundances of the \third $r$-process elements Os-Ir-Pt to iron
coincide with and exhibit a similar scatter to [Eu/Fe],  
again suggesting a similar origin for these four elements.
These new elemental abundance scatter data 
also appear to be consistent with the idea that not all supernovae will make 
Eu  (and Os-Pt) -- these appear to be rarer events 
than the synthesis of, for example, Ge and Zr -- and to 
point to  a 
lack of chemical homogeneity early in the history of the Galaxy.

The new abundance determinations for Os-Ir-Pt fall on the same
solar system scaled $r$-process curves as the rare-earth elements 
in the $r$-process rich stars \cs22 and \bd17. This agreement (or consistency)
now extends from Ba through the \third $r$-process peak for these stars,
again  indicating a similar synthesis origin 
for these elements.
The observed abundances of the 
lighter $n$-capture elements (Ge and Zr) do  not fall on the same solar
curve that matches the heavier such elements, 
and may indicate
two sites,  or at least astrophysical conditions, for the synthesis of 
all of the $n$-capture elements.

While the astrophysical site for the $r$-process has  still not
been precisely identified (see {\it e.g.}, Cowan \& Thielemann 2004), 
the abundance determinations presented here are consistent with 
a supernova origin and suggest that not all supernovae
may be responsible for synthesizing these $n$-capture elements. 
However, additional abundance determinations --particularly over a range of
mass number including both lighter and heavier $n$-capture elements -- 
in stars of
very low metallicity will be needed  to constrain models of $r$-process 
production, determine if there are multiple sites and to 
understand the history of element, and star,  formation at very early
times in the history of the Galaxy.

\acknowledgments

We thank Roberto Gallino,   Ken Nomoto and an anonymous referee 
for useful  discussions and comments that have helped us to improve the paper.
This work has been supported in part by NSF grants 
AST 03-07279 (J.J.C.),  AST 03-07495 (C.S.), 
AST 00-98508, AST 00-98549, AST 04-06784 (T.C.B.)  and AST 02-05124 (J.E.L.), 
and by STScI grants GO-8111 and GO-8342.  
Partial support was also provided 
by the NSF Frontier Center for Nuclear Astrophysics (JINA)
under grant PHY 02-16783 (J.W.T and T.C.B) 
and the DOE under grant DE-FG 02-91ER 40606 (J.W.T.).

\newpage
                                                                                
\appendix
                                                                                
\section{Substructure of Neutral Iridium Transitions}

Hyperfine and isotopic structure of \ion{Ir}{1} lines must be included 
in the synthesis of stellar spectra for accurate abundance determinations.   
Complete (reconstructed) line structure patterns of the three most important 
\ion{Ir}{1} lines are given in Table~\ref{tab5} for the users' convenience.  
Wavenumbers of transitions were taken from the improved energy levels 
(Ivarsson \etal\ 2003) and from Moore (1971) for the 3220~\AA\ line.  
The air wavelengths were computed from the energy levels using the 
standard index of air (Edl\'en 1953, 1966).  

Solar system isotopic abundances for Ir from Rosman \& Taylor (1998) were used. 
Both \iso{191}{Ir} and \iso{193}{Ir} have nuclear spin F~=~3/2.  
The first half of the listed components for each line are all from 
\iso{191}{Ir} and the second half are from \iso{193}{Ir}.  
The component strengths from \iso{191}{Ir} always sum to the solar system 
abundance of 0.37272 (within rounding error) and the component strengths 
from \iso{193}{Ir} sum to 0.62728 (within rounding error).   
Relative component strengths would need to be modified to model 
non-solar isotopic abundances.   
The isotope shift in the 3513~\AA\ line of --0.071~cm$^{\rm -1}$ for 
\iso{193}{Ir} with respect to \iso{191}{Ir} was adopted from Murakawa 
\& Suwa (1952), and that in the 3220~\AA\ and 3800~\AA\ lines of 
--0.00095~cm$^{\rm -1}$ and --0.064~cm$^{\rm -1}$, respectively,
were adopted from Sawatzky \& Winkler (1989).
Negative isotope shifts, where the heavier isotope is to the red of the 
lighter isotope, are common for heavy elements because the field shift 
from a non-negligible nuclear size overwhelms the normal and specific 
mass shifts.    

We took hyperfine A and B constants for the ground level from 
B\"uttgenbach \etal's (1978) extremely accurate atomic-beam 
magnetic-resonance measurements.
Upper level hyperfine constants were taken from Gianfrani \& Tito (1993)
for the 3220~\AA\ line and from Murakawa and Suwa (1952) for the
3513~\AA\ and 3800~\AA\ lines.
The log~$gf$~=~--0.52 for the 3220~\AA\ line was taken from 
Gough \etal\ (1983).
The transition probabilities listed by Ivarsson \etal\ (2003) were used to
compute log~$gf$~=~--1.21 for the 3513~\AA\ line and log~$gf$~=~--1.44 
for the 3800~\AA\ line.  
There is some inconsistency between the Einstein A coefficients in Table~4 
and the log~$gf$ values in Table~5 of Ivarsson \etal\ (2003).   
The log~$gf$ values in Table~5 of Ivarsson \etal\ are smaller than the 
above values by 0.04 or 0.05.  
Abundances in our work are based on the above log~$gf$ values
from Ivarsson \etal's Einstein~A coefficients which are correct 
(S. Ivarsson, private communication).
A similar problem was found in Ivarsson \etal's log(gf) for the Os I
3058~\AA\ line.  Our value of --0.40 is based on their A coefficient for this
line.


\tablenum{1}
\tablecolumns{7}
\tablewidth{0pt}
\label{tab1}

\begin{deluxetable}{lrrrccc}

\tabletypesize{\footnotesize}
\tablecaption{Stellar Model Parameters}
\tablehead{
\colhead{Star}                             &
\colhead{T$_{\rm eff}$}                    &
\colhead{log $g$}                          &
\colhead{v$_{\rm t}$}                      &
\colhead{[Fe I/H]}                         &
\colhead{[Fe II/H]}                        &
\colhead{References}                       }

\startdata
HD6268              & 4685 & 1.50 & 2.00 & --2.42 & --2.36 & 1 \\
HD6755              & 5105 & 2.95 & 2.50 & --1.68 & --1.57 & 2 \\
HD115444            & 4720 & 1.75 & 2.00 & --2.90 & --2.71 & 2 \\
HD122563            & 4570 & 1.35 & 2.90 & --2.72 & --2.61 & 2 \\
HD122956            & 4510 & 1.55 & 1.60 & --1.95 & --1.69 & 2 \\
HD126587            & 4795 & 1.95 & 2.00 & --2.93 & --2.81 & 1 \\
HD175305            & 5040 & 2.85 & 2.00 & --1.48 & --1.36 & 1 \\
HD186478            & 4600 & 1.45 & 2.00 & --2.56 & --2.44 & 2 \\
HD221170            & 4400 & 1.10 & 1.70 & --2.35 & --2.03 & 2 \\
BD+17$^{\rm o}$3248 & 5200 & 1.80 & 1.90 & --2.08 & --2.10 & 3 \\
CS 22892-052        & 4800 & 1.50 & 1.95 & --3.10 & --3.09 & 4 \\
\enddata

\tablenotetext{d}{The two Zr II lines at 3036~\AA\ form one blended feature.}

\tablerefs{(1) Derived in the same manner as the model atmospheres of
Simmerer \etal\ (2004); (2) Simmerer \etal\ (2004); (3) Cowan \etal\ (2002);
(4) Sneden \etal\ (2003).}

\end{deluxetable}

\tablenum{2}
\tablecolumns{16}
\tablewidth{0pt}
\label{tab2}

\begin{deluxetable}{lrrrrrrrrrrrrrr}

\rotate

\tabletypesize{\footnotesize}
\tablecaption{Abundances from Individual Transitions}
\tablehead{
\colhead{$\lambda$\tablenotemark{a}}       &
\colhead{Species}                          &
\colhead{E.P.\tablenotemark{b}}            &
\colhead{log $gf$}                         &
\colhead{(1)\tablenotemark{c}}             &
\colhead{(2)}                              &
\colhead{(3)}                              &
\colhead{(4)}                              &
\colhead{(5)}                              &
\colhead{(6)}                              &
\colhead{(7)}                              &
\colhead{(8)}                              &
\colhead{(9)}                              &
\colhead{(10)}                             &
\colhead{(11)}                             }

\startdata
3039.07                  & Ge I  & 0.88 & --0.04 &   +0.32 &   +1.08 &  --0.07 &    --0.16 &   +0.84 &     +0.03 &    +1.28 &   +0.35 &   +0.69 &   +0.46 & $<$--0.2 \\
3036.39\tablenotemark{d} & Zr II & 0.56 & --0.42 &   +0.31 &   +0.92 &  --0.13 &     +0.05 &   +0.96 &    --0.13 &    +1.17 &   +0.49 &   +1.08 &   +0.61 &  \nodata \\
3036.51\tablenotemark{d} & Zr II & 0.53 & --0.60 &         &         &         &           &         &           &          &         &         &         &          \\
3054.84                  & Zr II & 1.01 &  +0.18 &   +0.46 &   +1.02 &  --0.08 &   \nodata &   +1.2: &    --0.13 &    +1.32 &   +0.44 & \nodata &   +0.71 &    +0.15 \\
3060.12                  & Zr II & 0.04 & --1.37 &   +0.46 &   +1.07 &  --0.08 &     +0.10 &   +0.91 &  $<$--0.4 &    +1.17 &   +0.69 & \nodata &   +0.71 &    +0.30 \\
3061.33                  & Zr II & 0.10 & --1.38 &   +0.3: &   +1.02 &  --0.5: &     +0.03 &   +0.91 &  $<$--0.4 &    +1.22 &   +0.19 &   +0.68 &   +0.51 &    +0.50 \\
2838.63                  & Os I  & 0.64 &  +0.11 &  --0.5: &   +0.32 &  --0.60 &  $<$--0.8 &  --0.14 &    --0.6: &    +0.52 &  --0.6: &  --0.17 &   +0.45 &  \nodata \\
3058.65                  & Os I  & 0.00 & --0.40 &  --0.55: &   +0.45: & --0.75: &  $<$--1.2 &   -0.04 &  $<$--0.7 &    +0.67 &  --0.46 &  --0.15: &   +0.30 &    +0.05 \\
3301.57                  & Os I  & 0.00 & --0.74 & \nodata & \nodata &  --0.58 &  $<$--1.0 &  --0.09 &    --0.78 &    +0.32 &  --0.41 & \nodata &   +0.4: &    +0.02 \\
3220.78                  & Ir I  & 0.35 & --0.52 & \nodata & \nodata &  --0.88 &  $<$--1.3 &  --0.17 &    --1.08 &    +0.52 &  --0.71 & \nodata &   +0.25 &    +0.25 \\
3513.65                  & Ir I  & 0.00 & --1.21 & \nodata & \nodata &  --0.48 &  $<$--0.9 &  --0.02 &    --0.78 &    +0.47 &  --0.51 & \nodata &   +0.25 &   --0.10 \\
3800.12                  & Ir I  & 0.00 & --1.44 & \nodata & \nodata &  --0.78 &  $<$--1.3 &  --0.02 &    --1.08 &    +0.42 &  --0.71 & \nodata &   +0.20 &   --0.15 \\
2646.88                  & Pt I  & 0.00 & --0.79 &  --0.10 &   +0.7: & \nodata &   \nodata & \nodata &    --0.33 &    +1.07 &   +0.2: &   +0.98 &   +0.62 &  \nodata \\
2659.45                  & Pt I  & 0.00 & --0.03 &   +0.01 &   +0.20 &  --0.48 &  $<$--1.3 &   +0.3: &    --0.83 &    +0.37 &    0.0: &   +0.7: &   +0.42 &  \nodata \\
2929.79                  & Pt I  & 0.00 & --0.70 &   +0.01 &   +0.57 &  --0.48 &  $<$--1.3 &   +0.46 &    --0.78 &    +0.77 &  --0.16 &   +0.58 &   +0.52 &    +0.20 \\
\enddata

\tablenotetext{a}{Wavelength in~\AA.}

\tablenotetext{b}{Excitation potential in eV.}

\tablenotetext{c}{Star Names: (1)  HD6268; (2)  HD6755; (3)  HD115444; (4)  HD122563; (5)  HD122956; (6)  HD126587; \\
                              (7)  HD175305; (8)  HD186478; (9)  HD221170; (10) BD+17$^{o}$3248; (11) CS 22892-052.}

\tablenotetext{d}{The two Zr II lines at 3036~\AA\ form one blended feature.}

\end{deluxetable}

\tablenum{3}
\tablecolumns{9}
\tablewidth{0pt}
\label{tab3}

\begin{deluxetable}{lrrrrrrrc}

\tabletypesize{\footnotesize}
\tablecaption{Mean Abundances}
\tablehead{
\colhead{Star}                             &
\colhead{Ge}                               &
\colhead{Zr}                               &
\colhead{Os}                               &
\colhead{Ir}                               &
\colhead{Pt}                               &
\colhead{La}                               &
\colhead{Eu}                               &
\colhead{References}                       }

\startdata
\cutinhead{Abundances in log $\epsilon$ units}
Sun                 &    +3.60 &    +2.60 &    +1.41 &    +1.36 &    +1.70 &    +1.16 &    +0.52 & 1 \\
HD6268              &    +0.32 &    +0.38 &   --0.53: &  \nodata &   --0.01 &   --0.93 &   --1.33 & 2 \\
HD6755              &    +1.08 &    +1.01 &    +0.4: &  \nodata &    +0.45 &   --0.29 &   --0.50 & 3 \\
HD115444            &   --0.07 &   --0.20 &   --0.65 &   --0.68 &   --0.48 &   --1.35 &   --1.61 & 3 \\
HD122563            &   --0.16 &    +0.06 & $<$--1.2 & $<$--1.3 & $<$--1.3 &   --2.35 &   --2.59 & 3 \\
HD122956            &    +0.84 &    +1.00 &   --0.08 &    -0.05 &    +0.38 &   --0.48 &   --0.79 & 3 \\
HD126587            &    +0.03 &   \nodata&   --0.71:&   --0.96 &   --0.71 &   --1.70 &   --1.89 & 6 \\
HD175305            &    +1.28 &    +1.22 &    +0.50 &    +0.46 &    +0.67 &   --0.06 &   --0.29 & 6 \\
HD186478            &    +0.35 &    +0.45 &   --0.43 &   --0.63 &   --0.02 &   --1.29 &   --1.50 & 3 \\
HD221170            &    +0.69 &    +0.88 &   --0.16 &  \nodata &    +0.48 &   --0.72 &   --0.85 & 3 \\
BD+17$^{\rm o}$3248 &    +0.46 &    +0.64 &    +0.37 &    +0.23 &    +0.50 &   --0.42 &   --0.67 & 4 \\
CS 22892-052        & $<$--0.2 &    +0.32 &    +0.04 &   --0.05 &    +0.20 &   --0.84 &   --0.95 & 5 \\

\cutinhead{Abundances in [X/Fe] units}
Sun                 &     0.00 &     0.00 &     0.00 &     0.00 &     0.00 &     0.00 &     0.00 & 1 \\
HD6268              &   --0.86 &    +0.14 &    +0.49 &  \nodata &    +0.71 &    +0.26 &    +0.52 & 2 \\
HD6755              &   --0.84 &   --0.02 &    +0.67 &  \nodata &    +0.43 &    +0.12 &    +0.55 & 3 \\
HD115444            &   --0.77 &   --0.10 &    +0.84 &    +0.86 &    +0.72 &    +0.20 &    +0.58 & 3 \\
HD122563            &   --0.94 &    +0.07 &  $<$+0.1 &  $<$+0.1 & $<$--0.3 &   --0.90 &   --0.50 & 3 \\
HD122956            &   --0.81 &    +0.09 &    +0.46 &    +0.54 &    +0.63 &    +0.05 &    +0.38 & 3 \\
HD126587            &   --0.64 &   \nodata&    +0.81 &    +0.61 &    +0.52 &   --0.05 &    +0.40 & 6 \\
HD175305            &   --0.84 &   --0.02 &    +0.57 &    +0.58 &    +0.45 &    +0.14 &    +0.55 & 6 \\
HD186478            &   --0.69 &    +0.29 &    +0.68 &    +0.57 &    +0.84 &   --0.01 &    +0.42 & 3 \\
HD221170            &   --0.56 &    +0.31 &    +0.78 &  \nodata &    +1.13 &    +0.15 &    +0.66 & 3 \\
BD+17$^{\rm o}$3248 &   --1.06 &    +0.14 &    +1.04 &    +0.95 &    +0.88 &    +0.52 &    +0.91 & 4 \\
CS 22892-052        & $<$--0.7 &    +0.81 &    +1.73 &    +1.69 &    +1.60 &    +1.09 &    +1.62 & 5 \\
\enddata

\tablecomments{In the top section, La and Eu are taken from: 
(1) Lodders (2003); (2) Burris \etal\ (2000); (3) Simmerer \etal\ (2004); 
(4) Cowan \etal\ (2002); (5) Sneden \etal\ (2003); (6) this study.
In the bottom section the log~$\epsilon$ values derived from \ion{Ge}{1},
 \ion{Os}{1},  \ion{Ir}{1}, and \ion{Pt}{1} are referenced to the 
\ion{Fe}{1} values of Table~\ref{tab1}, while those from \ion{Zr}{2}, 
\ion{La}{2}, and \ion{Eu}{2} are referenced to \ion{Fe}{2}.}

\end{deluxetable}

\tablenum{4}
\tablecolumns{6}
\tablewidth{0pt}
\label{tab4}
                                                                                
\begin{deluxetable}{lccccc}
                                                                                
\tabletypesize{\footnotesize}
\tablecaption{Comparison of HST [Zr/Fe] Values}
\tablehead{
\colhead{Star}                             &
\colhead{HST}                              &
\colhead{Keck}                             &
\colhead{Burris}                           &
\colhead{Others}                           &
\colhead{References}                       }

\startdata
HD 6268      &  +0.14 & \nodata &   +0.32 &   +0.17,+0.12        & 1,2     \\
HD 6755      & --0.02 & \nodata &   +0.07 &   +0.08              & 3       \\
HD 115444    & --0.10 &   +0.05 & \nodata &   +0.18,+0.37,+0.26  & 2,4,5   \\
HD 122563    &  +0.07 &  --0.08 &   +0.31 &  --0.03,--0.08,+0.18,+0.03  
                                                                 & 2,3,4,5 \\
HD 122956    &  +0.09 &   +0.13 &   +0.16 &   +0.27              & 3       \\
HD 126587    & --0.09 &   +0.10 & \nodata &   +0.33,+0.12        & 2,5     \\
HD 175305    & --0.02 &   +0.14 &   +0.10 &   +0.16              & 3       \\
HD 186478    &  +0.29 &   +0.26 &   +0.40 &   +0.35,+0.28,+0.29  & 1,2,5   \\
HD 221170    &  +0.31 & \nodata &   +0.35 &   +0.12              & 3       \\
BD+17 3248   &  +0.14 &   +0.50 &   +0.27 &   +0.26,+0.25        & 5,6     \\
CS 22892-052 &  +0.81 &   +0.73 &   +0.79 &   +0.61,+0.60,+0.73  & 1,2,7   \\
\enddata
                                                                                
\tablecomments{References: (1) McWilliam et al. 1995;
                           (2) Honda et al. 2004;
                           (3) Fulbright 2000;
                           (4) Westin et al. 2000;
                           (5) Johnson 2002;
                           (6) Cowan et al. 2002;                               
                           (7) Sneden et al. 2003.}

\end{deluxetable}

\tablenum{5}
\tablecolumns{7}
\tablewidth{0pt}
\label{tab5}
                                                                                
\begin{deluxetable}{ccccrrc}
                                                                                
\tabletypesize{\footnotesize}
\tablecaption{Complete Line Structure Patterns for Three Ir I Transitions}
\tablehead{
\colhead{Frequency}                        &
\colhead{Wavelength}                       &
\colhead{Upper}                            &
\colhead{Lower}                            &
\colhead{Relative}                         &
\colhead{Relative}                         &
\colhead{Normalized}                       \\
\colhead{vacuum}                           &
\colhead{air}                              &
\colhead{F}                                &
\colhead{F}                                &
\colhead{Component}                        &
\colhead{Component}                        &
\colhead{Component}                        \\
\colhead{(cm$^{\rm -1}$)}                  &
\colhead{(\AA)}                            &
\colhead{}                                 &
\colhead{}                                 &
\colhead{Frequency}                        &
\colhead{Wavelength}                       &
\colhead{Strength}                         }
                                                                                
\startdata
31039.450 & 3220.7765 & 5 & 6 & --0.04119&  +0.00427 & 0.12113 \\
31039.450 & 3220.7765 & 5 & 5 &  +0.00578& --0.00060 & 0.00683 \\
31039.450 & 3220.7765 & 5 & 4 &  +0.06206& --0.00644 & 0.00016 \\
31039.450 & 3220.7765 & 4 & 5 & --0.01616&  +0.00168 & 0.09566 \\
31039.450 & 3220.7765 & 4 & 4 &  +0.04012& --0.00416 & 0.00897 \\
31039.450 & 3220.7765 & 4 & 3 &  +0.09635& --0.01000 & 0.00019 \\
31039.450 & 3220.7765 & 3 & 4 &  +0.02388& --0.00248 & 0.07474 \\
31039.450 & 3220.7765 & 3 & 3 &  +0.08011& --0.00831 & 0.00679 \\
31039.450 & 3220.7765 & 2 & 3 &  +0.06870& --0.00713 & 0.05824 \\
31039.450 & 3220.7765 & 5 & 6 & --0.04666&  +0.00484 & 0.20387 \\
31039.450 & 3220.7765 & 5 & 5 &  +0.00692& --0.00072 & 0.01150 \\
31039.450 & 3220.7765 & 5 & 4 &  +0.06707& --0.00696 & 0.00026 \\
31039.450 & 3220.7765 & 4 & 5 & --0.01662&  +0.00172 & 0.16100 \\
31039.450 & 3220.7765 & 4 & 4 &  +0.04353& --0.00452 & 0.01509 \\
31039.450 & 3220.7765 & 4 & 3 &  +0.10179& --0.01056 & 0.00033 \\
31039.450 & 3220.7765 & 3 & 4 &  +0.02573& --0.00267 & 0.12578 \\
31039.450 & 3220.7765 & 3 & 3 &  +0.08399& --0.00872 & 0.01143 \\
31039.450 & 3220.7765 & 2 & 3 &  +0.07123& --0.00739 & 0.09801 \\
          &           &   &   &          &           &         \\
28452.318 & 3513.6473 & 7 & 6 &  +0.1034 & --0.01277 & 0.11648 \\
28452.318 & 3513.6473 & 6 & 6 &  +0.0348 & --0.00430 & 0.00459 \\
28452.318 & 3513.6473 & 6 & 5 &  +0.0568 & --0.00701 & 0.09636 \\
28452.318 & 3513.6473 & 5 & 6 & --0.0167 &  +0.00207 & 0.00007 \\
28452.318 & 3513.6473 & 5 & 5 &  +0.0053 & --0.00065 & 0.00606 \\
28452.318 & 3513.6473 & 5 & 4 &  +0.0116 & --0.00143 & 0.07929 \\
28452.318 & 3513.6473 & 4 & 5 & --0.0326 &  +0.00402 & 0.00008 \\
28452.318 & 3513.6473 & 4 & 4 & --0.0263 &  +0.00324 & 0.00457 \\
28452.318 & 3513.6473 & 4 & 3 & --0.0291 &  +0.00359 & 0.06523 \\
28452.318 & 3513.6473 & 7 & 6 &  +0.0379 & --0.00468 & 0.19602 \\
28452.318 & 3513.6473 & 6 & 6 & --0.0356 &  +0.00440 & 0.00772 \\
28452.318 & 3513.6473 & 6 & 5 & --0.0136 &  +0.00168 & 0.16217 \\
28452.318 & 3513.6473 & 5 & 6 & --0.0921 &  +0.01137 & 0.00012 \\
28452.318 & 3513.6473 & 5 & 5 & --0.0700 &  +0.00865 & 0.01019 \\
28452.318 & 3513.6473 & 5 & 4 & --0.0626 &  +0.00773 & 0.13344 \\
28452.318 & 3513.6473 & 4 & 5 & --0.1125 &  +0.01389 & 0.00014 \\
28452.318 & 3513.6473 & 4 & 4 & --0.1050 &  +0.01296 & 0.00770 \\
28452.318 & 3513.6473 & 4 & 3 & --0.1061 &  +0.01310 & 0.10977 \\
          &           &   &   &          &           &         \\
26307.462 & 3800.1239 & 6 & 6 &  +0.0970 & --0.01401 & 0.11563 \\
26307.462 & 3800.1239 & 6 & 5 &  +0.1190 & --0.01718 & 0.00551 \\
26307.462 & 3800.1239 & 5 & 6 &  +0.0248 & --0.00358 & 0.00551 \\
26307.462 & 3800.1239 & 5 & 5 &  +0.0468 & --0.00676 & 0.08976 \\
26307.462 & 3800.1239 & 5 & 4 &  +0.0531 & --0.00767 & 0.00723 \\
26307.462 & 3800.1239 & 4 & 5 & --0.0013 &  +0.00019 & 0.00723 \\
26307.462 & 3800.1239 & 4 & 4 &  +0.0050 & --0.00072 & 0.07120 \\
26307.462 & 3800.1239 & 4 & 3 &  +0.0022 & --0.00031 & 0.00544 \\
26307.462 & 3800.1239 & 3 & 4 & --0.0256 &  +0.00370 & 0.00544 \\
26307.462 & 3800.1239 & 3 & 3 & --0.0284 &  +0.00411 & 0.05979 \\
26307.462 & 3800.1239 & 6 & 6 &  +0.0347 & --0.00502 & 0.19460 \\
26307.462 & 3800.1239 & 6 & 5 &  +0.0567 & --0.00820 & 0.00927 \\
26307.462 & 3800.1239 & 5 & 6 & --0.0396 &  +0.00572 & 0.00927 \\
26307.462 & 3800.1239 & 5 & 5 & --0.0176 &  +0.00254 & 0.15107 \\
26307.462 & 3800.1239 & 5 & 4 & --0.0101 &  +0.00146 & 0.01217 \\
26307.462 & 3800.1239 & 4 & 5 & --0.0686 &  +0.00991 & 0.01217 \\
26307.462 & 3800.1239 & 4 & 4 & --0.0611 &  +0.00883 & 0.11982 \\
26307.462 & 3800.1239 & 4 & 3 & --0.0622 &  +0.00899 & 0.00915 \\
26307.462 & 3800.1239 & 3 & 4 & --0.0948 &  +0.01369 & 0.00915 \\
26307.462 & 3800.1239 & 3 & 3 & --0.0959 &  +0.01386 & 0.10063 \\
\enddata
                                                                                
\end{deluxetable}

\clearpage

\clearpage

\clearpage

\clearpage

\clearpage

\clearpage

\clearpage


\begin{thebibliography}{}


\bibitem[]{aam99}
Alonso, A., Arribas, S., \& Martinez-Roger, C. 1999, \aaps, 140, 261
                                                                                
\bibitem[]{aam01}
Alonso, A., Arribas, S., \& Martinez-Roger, C. 2001, \aap, 376, 1039

\bibitem[]{twa94} 
Anthony-Twarog, B.~J.~\& Twarog, B.~A.\ 1994, \aj, 107, 1577

\bibitem[]{arg04}
Argast, D., Samland, M., Thielemann, F.-K., \& Qian, Y.-Z 2004,
\aap, 416, 997 

\bibitem[]{bar97}
Barlow, T. A. \& Sargent, W. L. W. 1997, \aj, 113, 136

\bibitem[]{bie81} 
Bi\'emont, E.,  
Grevesse, N., Hannaford, P., \& Lowe, R. M., 1981, ApJ, 248, 867 

\bibitem[]{bie99} 
Bi\'emont, E., Lyng{\aa}, C., Li, Z. S., Svanberg, S., Garnir, H. P.,
\& Doidge, P. S. 1999, \mnras, 303, 721

\bibitem{bur00}
Burris, D. L., Pilachowski, C. A., Armandroff, T. A., Sneden, C.,
Cowan, J. J., \& Roe, H. 2000, \apj, 544, 302

\bibitem{but78}
B\"uttgenbach, S., Dicke, R., Gebauer, H., Kuhnen, R., \& Tr\"aber, F. 
1978, Z. Phys. A, 286, 333

\bibitem{cam01}
Cameron, A. G. W. 2001, \apj, 562, 456

\bibitem[]{cas97}
Castelli, F., Gratton, R. G., \& Kurucz, R. L. 1997, \aap, 318, 841

\bibitem{cay04}
Cayrel, R., et al. 2004, A\&A, 416, 1117

\bibitem{chi04}
Chieffii, A.,  \& Limongi, M.  2004, ApJ, 608, 405

\bibitem{chr04}
Christlieb, N., et al. 2004, Nature, 603, 708

\bibitem[Corliss & Bozman 1962]{CB62}
Corliss, C. H. \& Bozman, W. R. 1962, Experimental Transition
Probabilities for Spectral Lines of Seventy Elements, NBS Monograph 53
(Washington: US Gov. Prt. Off.)

\bibitem{cow02}
Cowan, J. J., \etal\  2002, \apj, 572, 861

\bibitem{cow95}
Cowan, J. J.,  Sneden, C., Burris, D. L., \& Truran, J. W. 
1996, \apj, 460, L115

\bibitem[Cowan \& Sneden 2004]{cow04a}
Cowan, J. J., \& Sneden, C. 2004, in Carnegie Observatories Astrophysics 
Series, Vol. 4: Origin and Evolution of the Elements, ed. A. McWilliam 
\& M. Rauch (Cambridge: Cambridge Univ. Press), in press

\bibitem[]{cow04b}
Cowan, J. J., \& Thielemann, F.-K., 2004, Phys. Today, in press

\bibitem{del73}
Delbouille, L, Roland, G., \& Neven, L. 1973, Photometric Atlas of the
Solar Spectrum from lambda 3000 to lambda 10000, (Li`ege, Inst. d'Ap.,
Univ. de Li`ege)

\bibitem{den03}
Den Hartog, E. A., Lawler, J. E., Sneden, C., \& Cowan, J. J. 2003, 
\apjs,  148, 543   

\bibitem{den05}
Den Hartog, E. A.,  Herd, T. M., Lawler, J. E., Sneden, C., 
Cowan, J. J., \& Beers, T. C. 2005, \apj, in press 

\bibitem{edl53}
Edl\'en, B. 1953, J. Opt. Soc. Am., 43, 339

\bibitem{edl66}
Edl\'en, B. 1966, Metrologia, 2, 71

\bibitem{field02}
Fields, B. D., Truran, J. W., \& Cowan, J. J. 2002, \apj,  575, 845

\bibitem{frb04}
Frebel, A., et al. 2004, Nature, submitted

\bibitem{fre99}
Freiburghaus, C., Rosswog, S., \& Thielemann, F.-K. 1999, \apj, 525, L121

\bibitem{ful00} Fulbright, J. P. 2000, \aj, 120, 1841

\bibitem{gia93}
Gianfrani, L. \& Tito G. M. 1993, Z. Phys. D, 25, 113

\bibitem{gil88}
Gilroy, K. K., Sneden, C., Pilachowski, C. A.,  \&  Cowan, J. J. 1988,
\apj, 327, 298

\bibitem{gou83} 
Gough, D. S., Hannaford, P., \& Lowe, R. M. 1983, J. Phys. B., 16, 785

\bibitem{gra99} Gratton, R. G., Carretta, E., Eriksson, K., \& Gustafsson, B. 
1999, \aap, 350, 955

\bibitem{heg02}
Heger, A., \& Woosley, S. E. 2002, ApJ, 567, 532

\bibitem{hil02}
Hill, V., \etal\ 2002, \aap, 387, 560

\bibitem{hof01}
Hoffman, R. D., Woosley, S. E., \& Weaver, T. A.  2001, 
ApJ, 549, 1085

\bibitem{hon04}
Honda, S., Aoki, W, Kajino, T., Ando, H., Beers, T. C.,
Izumiura, H., Sadakane, K., \& Takada-Hidai, M. 2004,
\apj,  607, 474

\bibitem{iva03}
Ivarsson, S., \etal\ 2003, \aap, 409, 1141

\bibitem{joh02} Johnson, J. A. 2002, \apjs, 139, 219 

\bibitem{kor04} Korn, A. J. 2004, Origin and Evolution of the Elements,
Carnegie Observatories Centennial Symposia, Carnegie Observatories 
Astrophysics Series, ed. A. McWilliam and M. Rauch,
$\rm http://www.ociw.edu/ociw/symposia/series/symposium4/proceedings.html$

\bibitem[]{kur95}
Kurucz, R. L. 1995, in Workshop on Laboratory and astronomical high
resolution spectra, ASP Conference Ser. \#81 ed. A.J. Sauval, R. Blomme,
and N. Grevesse (San Francisco: Astr. Soc. Pac.), p.583

\bibitem{kwi84}
Kwiatkowski, M., Zimmermann, P., Bi\'emont, E., \& Grevesse, N.
1984, \aap, 135, 59

\bibitem{law04}
Lawler, J. E., Sneden, C., \& Cowan, J. J. 2004, \apj, 608, 850 

\bibitem{lod03}
Lodders, K. 2003, ApJ,  591, 1220

\bibitem{mcw95} McWilliam, A., Preston, G. W., Sneden, C., \& Searle, L. 
1995, \aj, 109, 2757 

\bibitem{moo71}
Moore, C. E. 1971, Atomic Energy Levels: As Derived from the Analysis of
Optical Spectra, Vol. 3, National Standard Reference Data Series-National
Bureau of Standards 35  (Washington, DC: U. S. Gov. Printing Off.), 177

\bibitem{mur52}
Murakawa, K. \& Suwa S. 1952, Phys. Rev. 87, 1048

\bibitem{per97} 
Perryman, M.~A.~C., et~al.\ 1997, \aap, 323, L49

\bibitem{ros98}
Rosman, K. J. R., \& Taylor, P. D. P. 1998, J. Phys. Chem. Ref. Data 27, 1275

\bibitem{ros99}
Rosswog, S., Liebendorfer, M., Thielemann, F.-K., Davies, M. B., Benz, W.,
\& Piran, T., 1999, \aap, 341, 499

\bibitem{saw89}
Sawatzky, G.,  \& Winkler, R.  1989, Z. Phys. D, 14, 9 

\bibitem{sim04}
Simmerer, J., Sneden, C.,  Cowan, J. J., Collier, J.,  Woolf, V., 
\& Lawler, J. E. 2004, \apj, in press 

\bibitem{sne98} 
Sneden, C., Cowan, J. J., Burris, D. L., \& Truran, J. W. 1998, \apj, 496, 235

\bibitem{sne03}
Sneden, C.,  \etal\ 2003, \apj, 591, 936

\bibitem{sne03a}
Sneden, C., \& Cowan, J. J. 2003, Science, 299, 70

\bibitem{sne00}
Sneden, C., Cowan, J. J., Ivans, I. I., Fuller, G. M., Burles, S.,
Beers, T. C., \& Lawler, J. E.  2000, \apj, 533, L139


\bibitem{sne96}
Sneden, C., McWilliam, A., Preston, G. W., Cowan, J. J., Burris, D. L.,
\& Armosky, B. J. 1996, \apj, 467, 819

\bibitem[]{the99} 
Th{\' e}venin, F.~\& Idiart, T.~P.\ 1999, \apj, 521, 753

\bibitem{tra04}
Travaglio, C.,  Gallino, R.,  Arnone, E., Cowan, J. J., Jordan, F.,
\& Sneden, C.  2004, \apj,  601, 864

\bibitem{tru02}
Truran, J. W., Cowan, J. J., Pilachowski, C. A., \& Sneden, C.
2002, \pasp,  114, 1293

\bibitem[]{ume04}
Umeda, H., \& Nomoto, K. 2005, \apj, in press, astro-ph/0308029

\bibitem[]{Vetal94}
Vogt, S. S. \etal\ 1994, in Proc. SPIE Conf. 2198,
Instrumentation in Astronomy VII, eds. D. L.  Crawford and E. R. Craine,
(Bellingham, WA: SPIE), p. 362

\bibitem{was96}
Wasserburg, G. J., Busso, M., \& Gallino, R. 1996, \apj, 466, L109

\bibitem{was00}
Wasserburg, G. J.,  \&  Qian, Y.-Z. 2000, \apj,  529, L21


\end{thebibliography}
\end{document}